\documentclass[aps,amsmath,amssymb,prl,twocolumn,longbibliography,floatfix]{revtex4-2}

\usepackage{mathrsfs} 
\usepackage{graphicx,color} 
\usepackage{mathtools}
\usepackage{threeparttable}
\usepackage{bigints}
\usepackage{bm}
\usepackage[normalem]{ulem}
\usepackage{physics}
\usepackage{color}
\usepackage[colorlinks=true]{hyperref}
\hypersetup{
  colorlinks=true,
  linkcolor=blue,
  filecolor=magenta,
  citecolor=blue,
  urlcolor=blue,
}
\newcommand{\av}[1]{\langle{#1}\rangle}

\newcommand{\smallsquare}{\scalebox{0.6}{$\square$}}

\begin{document}

\title{Elastic heterogeneity governs anomalous scaling in a soft porous crystal}%72

\author{Kota Mitsumoto}
%\email{kmitsu@iis.u-tokyo.ac.jp}
\email{kmitsumoto@g.ecc.u-tokyo.ac.jp}
\affiliation{Graduate School of Arts and Sciences, The University of Tokyo, 3-8-1 Komaba, Meguro-ku, Tokyo 153-8902, Japan}
\author{Kyohei Takae}%
%\email{takae@iis.u-tokyo.ac.jp}
\email{takae@tottori-u.ac.jp}
%\affiliation{Department of Fundamental Engineering, Institute of Industrial Science, University of Tokyo, 4-6-1 Komaba, Meguro-ku, Tokyo 153-8505, Japan}
\affiliation{Graduate School of Engineering, Tottori University, 4-101 Koyama-Cho Minami, Tottori, Tottori 680-8552, Japan}
%\affiliation{Advanced Mechanical and Electronic System Research Center, Faculty of Engineering, Tottori University, 4-101 Koyama-cho-minami, Tottori, Tottori 680-8552, Japan}

\date{\today}

\begin{abstract}
Nanoscale molecular transport plays a crucial role in regulating mass diffusion and responsiveness in condensed matter systems~\cite{Mehrer}. In soft porous crystals~\cite{horike2009soft}, in particular, adsorption of guest molecules induces host framework deformation and changes in rigidity, underpinning their characteristic stimuli-responsive behaviour~\cite{gor2017adsorption,coudert2025recent}. Surface-mediated adsorption leads to inhomogeneous adsorbate distribution~\cite{Karger}, which, through local framework deformation, induces spatial variations in rigidity---elastic heterogeneity~\cite{Onukibook}. Although this heterogeneity is expected to affect adsorption kinetics and mechanical behaviour, its role remains poorly understood. Here we show that elastic heterogeneity governs adsorption kinetics, leading to emergent phenomena including size-dependent uptake, surface creasing, and anomalous dynamic scaling that is distinct from established scaling. Stress relaxation near corners facilitates adsorption, resulting in a size-dependent deviation from diffusive kinetics. Away from corners, flexible unadsorbed regions between rigid adsorbed domains relieve stress through crease formation. The resulting lateral correlations exhibit anomalous dynamic scaling, characterized by a breakdown of scale invariance between global and local interfacial fluctuations~\cite{ramasco2000generic}. These findings provide a mechanistic foundation for controlling adsorption and deformation kinetics in soft porous materials via elastic heterogeneity. Our work opens a route to engineering responsive materials, where mechanical feedback is harnessed to control cooperative molecular transport and drive macroscopic shape changes under external perturbations.
\end{abstract}

\maketitle

%%%%%%%%%%%%%%%%%%%%%%%%%%%%%%%%%%%%%%%%
%%%%%%%%%%%%%%%%%%%%%%%%%%%%%%%%%%%%%%%%
% introduction
%%%%%%%%%%%%%%%%%%%%%%%%%%%%%%%%%%%%%%%%
%%%%%%%%%%%%%%%%%%%%%%%%%%%%%%%%%%%%%%%%

\noindent
Transport phenomena govern the structure, dynamics, and function of systems spanning physics, chemistry, biology, and materials science. Among these, mass transport plays a central role in processes ranging from droplet formation in biological cells~\cite{lee2021chromatin}, ionic conduction in batteries~\cite{mukhopadhyay2014deformation,yabuuchi2014research}, hydrogen uptake in metals~\cite{alefeld1978hydrogen}, and solvent permeation in gels and soils~\cite{li1992phase,doi2009gel,Coussy2011}, to molecular diffusion in porous materials such as metal–organic frameworks (MOFs)~\cite{mitra2021diffusion,horike2009soft,horike2013ion,furukawa2015heterogeneity,lim2020proton}. Designing materials with tunable responsiveness to external stimuli requires a fundamental understanding of the material parameters that regulate molecular transport~\cite{krishna2012diffusion,Karger,coudert2025recent}. Recent advances in quantum chemistry and molecular dynamics simulations have elucidated how host–guest interactions influence adsorption energetics and the dynamic response of the framework~\cite{odoh2015quantum,jawahery2017adsorbate,rogge2019unraveling,schaper2023simulating}. However, although these approaches provide microscopic insight, building predictive theoretical frameworks that bridge molecular-scale mechanisms and macroscopic kinetics remains a major challenge. Cooperative transport among guest molecules, combined with local structural flexibility, gives rise to spatio-temporal heterogeneity in material properties, making the system intrinsically non-uniform. This non-uniformity leads to heterogeneous and size-dependent adsorption through surface-controlled uptake, as extensively observed in MOFs experimentally and numerically~\cite{sakata2013shape,karger2014microimaging,cho2015extra,sakaida2016crystalline,hosono2019highly,rogge2019unraveling,delen2021in,schaper2023simulating,yamada2024three,watanabe2024size}. Such adsorption behaviour leads to complex spatio-temporal dynamics and presents a major challenge for predicting and controlling uptake processes. Establishing physical principles to address such mesoscopic heterogeneity is therefore essential for engineering MOF-based devices with enhanced functionalities, including water harvesting~\cite{hanikel2021evolution}, catalysis~\cite{bavykina2020metal}, sensors~\cite{kreno2012metal}, biomedicines~\cite{horcajada2012metal}, and artificial molecular machines~\cite{danowski2019unidirectional}.

One major origin of adsorption heterogeneity arises from the coupling between molecular uptake and the elastic responses of the host framework~\cite{horike2009soft,gor2017adsorption}. In soft porous crystals---mechanically flexible MOFs---strong host–guest interactions lead not only to substantial structural deformation but also to spatial variations in rigidity, referred to as elastic heterogeneity~\cite{eshelby1957determination,Onukibook,mitsumoto2023elastic}. This mechanical heterogeneity gives rise to cooperative effects, which manifest as pattern formation, including domain formation and superlattice ordering, mediated by long-range elastic interactions in the host framework~\cite{cho2015extra,jawahery2017adsorbate,mitsumoto2024adsorption}. However, the impact of elastic heterogeneity on adsorption kinetics, in both bulk and near-surface regions, remains poorly understood.

\begin{figure*}[t]
\centering
\includegraphics[width=175mm]{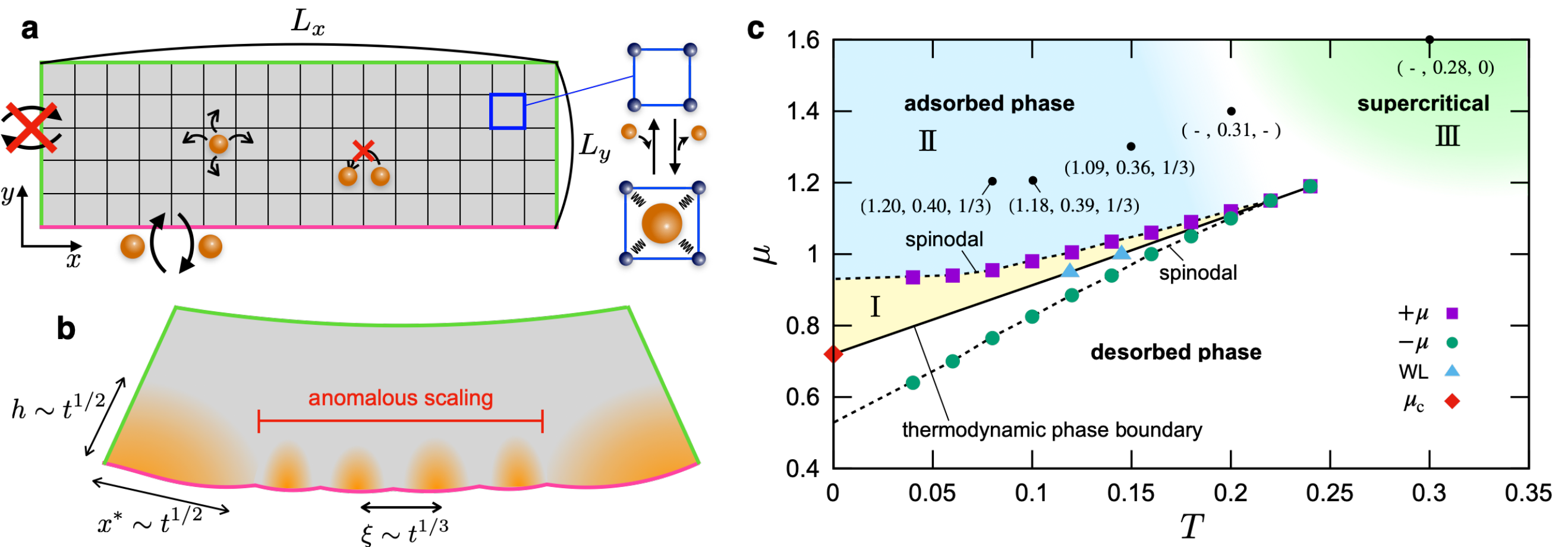}
\caption{
{\bf Schematic description of the numerical simulation.}
{\bf a,} The setup of molecular adsorption kinetics. Guest particles, represented by orange spheres, enter from the bottom boundary ($y=0$), indicated by the magenta line. Lattice expansion and hardening occur due to interactions between the host matrix and guest particles. The other boundaries, represented by the green lines, do not allow the entrance and exit of particles, whereas stress relaxation occurs to satisfy stress-free boundary conditions. Inside the host matrix, guest particles move under the influence of other particles and the host's elasticity. The interparticle interaction is governed by simple exclusion, where each matrix site can adsorb at most one guest particle.
{\bf b,} Pattern formation and elastic creasing during adsorption in the spinodal region. Adsorbed (desorbed) regions are displayed in orange (gray). Molecular adsorption proceeds faster at the two bottom corners, whereas adsorption domains appear at the bottom surface. The domains at the corners grow both laterally and vertically as $x^*\sim t^{1/2}$ and $h\sim t^{1/2}$, respectively (see Fig.~\ref{fig::distribution}b). Unadsorbed regions between domains exhibit creasing to relax mechanical stresses with the correlation length $\xi\sim t^{1/3}$. The domain morphology satisfies an anomalous dynamic scaling relation, as shown in Fig.~\ref{fig::scaling}.
{\bf c,} Thermodynamic phase diagram for $(K,\lambda)=(3,1.4)$ with respect to temperature $T$ and adsorption chemical potential $\mu$. The solid line represents the equilibrium phase boundary between the adsorbed and desorbed phases. The dashed lines represent the transition points under quasi-equilibrium simulations with increasing $\mu$ ($+\mu$) and decreasing $\mu$ ($-\mu$), exhibiting hysteretic behaviour. Dynamic scaling exponents ($\alpha, \beta, 1/z$) for several ($T, \mu$) are also displayed (see Table~\ref{tab::scaling}).
}
\label{fig::summary}
\end{figure*}

To uncover the underlying principles governing heterogeneous adsorption kinetics, it is crucial to determine whether these dynamics obey universal scaling laws. Scaling concepts from interfacial growth phenomena~\cite{krug1991kinetic,Barabasi,takeuchi2012evidence}---such as swelling-deswelling in gels~\cite{tanaka1987mechanical,matsuo1992patterns,Onukibook,doi2009gel} and epitaxial thin-film growth~\cite{politi2000instabilities,aqua2013growth,Onukibook}---have revealed self-affine evolutions governed by elasticity and interfacial tension. Given that adsorption in soft porous crystals involves surface-mediated transport coupled to elastic stress, it is natural to explore whether similar scaling descriptions apply. Establishing such scaling laws would provide a unified framework for understanding how elastic heterogeneity and the resulting cooperative guest transport govern adsorption kinetics.

\begin{figure*}[t]
\centering
\includegraphics[width=175mm]{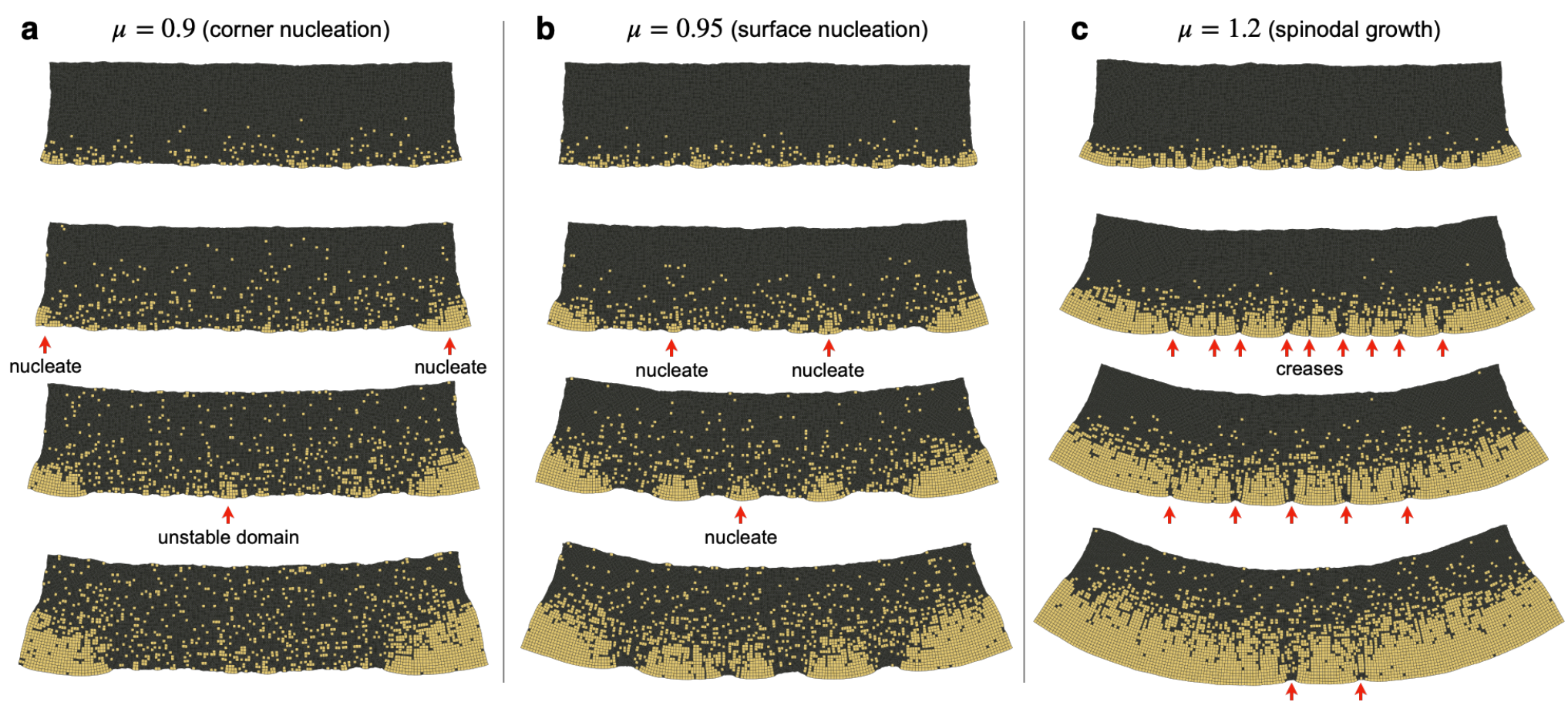}
\caption{
{\bf Nucleation and spinodal growth of molecular adsorption.}
{\bf a,} When the desorbed state is metastable, domain nucleation of adsorbates occurs at the corners, while adsorbed domains do not grow at the surface (see Supplementary Video 1).
{\bf b,} When the chemical potential approaches the spinodal point, surface nucleation also occurs (see Supplementary Video 2).
{\bf c,} When the desorbed state is unstable, molecular adsorption exhibits a spinodal-type growth (see Supplementary Video 3). Adsorbed domains grow with a characteristic wavenumber. Because unadsorbed regions between adsorbed domains are softer, creases appear in the unadsorbed regions to reduce elastic energy, as indicated by red arrows. The parameters for all panels are: $L_x=192$, $K=3.0$, $\lambda=1.4$, and $T=0.1$.
}
\label{fig::snapshot}
\end{figure*}

%%%%%%%%%%%%%%%%%%%%%%%%%%%%%%%%%%%%%%%%
%%%%%%%%%%%%%%%%%%%%%%%%%%%%%%%%%%%%%%%%
% thermodynamics
%%%%%%%%%%%%%%%%%%%%%%%%%%%%%%%%%%%%%%%%
%%%%%%%%%%%%%%%%%%%%%%%%%%%%%%%%%%%%%%%%
%\subsection*{Results}

%fig 1a
However, to extract scaling exponents, it is necessary to perform large-scale simulations of adsorption kinetics with statistically converged ensemble averaging. Because soft porous crystals consist of metallic nodes and organic linkers, molecular dynamics simulations involving many unit cells remain computationally demanding~\cite{jawahery2017adsorbate,rogge2019unraveling,schaper2023simulating}.

As a tractable alternative, we develop a coarse-grained lattice model that captures the elastic deformation of soft porous crystals, as schematically illustrated in Fig.~\ref{fig::summary}a (see Methods for details). Each lattice site can adsorb at most one guest particle, and particles hop to adjacent sites subject to excluded volume constraints. The model includes two key material parameters: the rigidity change factor $K$, and the lattice expansion ratio $\lambda$. Specifically, $K>1$ and $\lambda>1$ correspond to local lattice hardening and expansion upon adsorption, respectively. In this study, we mainly focus on the representative case of expansion and hardening, $(K,\lambda)=(3,1.4)$, where heterogeneous adsorption kinetics becomes pronounced. Adsorption occurs only from the bottom boundary, mimicking directional uptake. These simplifications allow us to extract scaling behaviour with a feasible numerical cost. We use dynamic Monte Carlo simulations to examine the adsorption and diffusion kinetics (see Methods for details).

%fig1b
Before presenting the details of the numerical results, we highlight a representative pattern formation, as schematically displayed in Fig.~\ref{fig::summary}b. Two major adsorption domains form at the bottom corners, whereas smaller domains appear along the surface away from the corners. Since molecular adsorption induces lattice expansion and proceeds from the bottom, the system undergoes macroscopic bending to relax elastic stress. Notably, adsorption domains are separated by unadsorbed regions. Unadsorbed softer regions deform locally, forming elastic creases. In contrast, deformation in the adsorbed domains remains small due to local hardening. As described later, the adsorption distribution exhibits a characteristic time evolution known as anomalous scaling. The snapshot in Fig.~\ref{fig::summary}b corresponds to the case with lattice expansion and hardening. Under different parameter conditions, qualitatively distinct patterns are observed (see Extended Data Fig.~\ref{fig::ex::schematic}a--c).

\begin{figure*}[t]
\centering
\includegraphics[width=175mm]{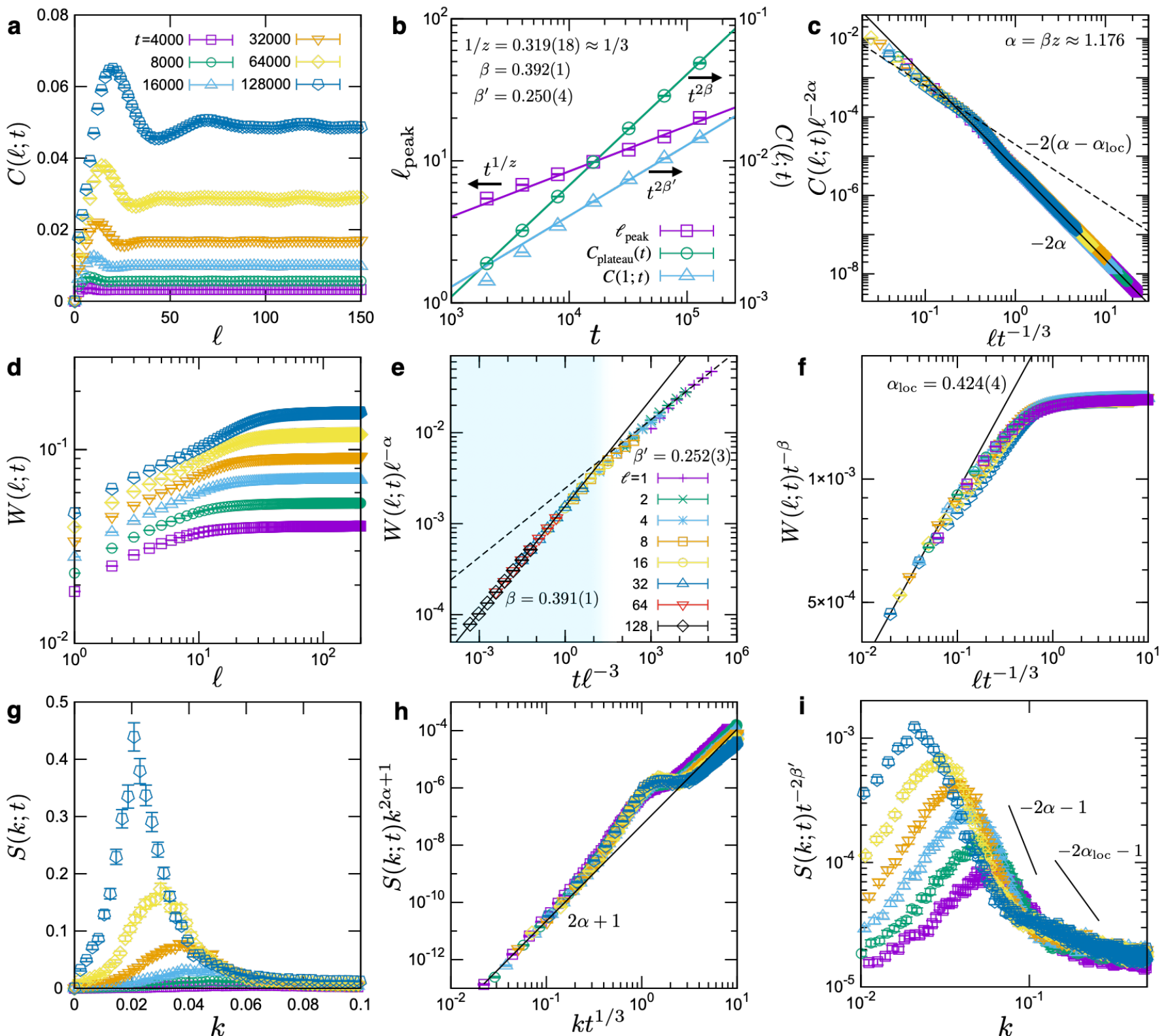}
\caption{
{\bf Dynamic scaling of adsorption distribution.}
{\bf a,} Spatial correlation function $C(\ell;t)$ exhibits a maximum at $\ell_{\rm peak}(t)$ and a plateau $C_{\rm plateau}(t)$ for large $\ell$.
{\bf b,} Time evolution of $\ell_{\rm peak}(t)$, $C_{\rm plateau}(t)$, and $C(1;t)$. They exhibit power-law behaviours, yielding the dynamic exponent $1/z=0.319(18) \simeq 1/3$, growth exponent $\beta=0.392(1)$, and anomalous growth exponent $\beta'=0.250(4)$, respectively. The positive $\beta'$ indicates anomalous dynamic scaling. Data in the interval $[8000,128000]$ are used for the fitting.
{\bf c,} Scaling plot $\tilde{C}(\ell t^{-1/3})=C(\ell;t)\ell^{-2\alpha}$, where $\alpha=\beta z$ represents the global roughness exponent. 
{\bf d,} Adsorption deviation $W(\ell;t)$.
{\bf e,} Scaling plot $\tilde{W}_1(t\ell^{-3})=W(\ell;t)\ell^{-\alpha}$ exhibits a crossover between distinct slopes $\beta=0.391(1)$ and $\beta'=0.252(3)$ at $t\ell^{-3}\simeq 24.4$.
{\bf f,} Another scaling plot $\tilde{W}_2(\ell t^{-1/3})=W(\ell;t)t^{-\beta}$. Data collapse for $\ell t^{-1/3}\lesssim 0.1$ yields the local roughness exponent $\alpha_{\rm loc}=0.424(4)$, which corresponds to $(\beta-\beta')z$.
{\bf g,} Structure factor $S(k;t)$. 
{\bf h,} Scaling plot $\tilde{S}(k t^{1/3})=S(k;t)k^{2\alpha+1}$ using the global roughness exponent $\alpha$ collapses for $k t^{1/3}\lesssim 1$, whereas a plateau is observed at $k t^{1/3}\sim 1$.
{\bf i,} Another scaling plot $\tilde{S}(k)=S(k;t)t^{-2\beta'}$ using the anomalous growth exponent $\beta'$. The slope reads $k^{-2\alpha-1}$ for $k\lesssim 10^{-1}$, whereas deviation from both $k^{-2\alpha-1}$ and $k^{-2\alpha_{\rm loc}-1}$ is confirmed for $k\gtrsim 10^{-1}$, though there is a good data collapse. The parameters for all panels are: $L_x=768$, $K=3.0$, $\lambda=1.4$, $T=0.1$, and $\mu=1.2$.
}
\label{fig::scaling}
\end{figure*}

%fig1c
The adsorption distribution in Fig.~\ref{fig::summary}b is observed above the spinodal curve of the phase diagram, which is shown in Fig.~\ref{fig::summary}c (see Methods for numerical details to obtain the phase diagram). The adsorption kinetics depend sensitively on temperature $T$ and adsorption chemical potential $\mu$ as well as material parameters. In the phase diagram, the hysteretic adsorption transition is a consequence of the attractive interaction between adsorbed molecules, which arises from lattice deformation upon adsorption (see Methods for the Landau theory). As described later, anomalous adsorption kinetics at the surface are classified using a dynamic scaling ansatz. The obtained scaling exponents for selected $T$ and $\mu$ are displayed in Fig.~\ref{fig::summary}c (see Fig.~\ref{fig::scaling} and Table~\ref{tab::scaling} for details).

%%%%%%%%%%%%%%%%%%%%%%%%%%%%%%%%%%%%%%%%
%%%%%%%%%%%%%%%%%%%%%%%%%%%%%%%%%%%%%%%%
% surface and corner contribution
%%%%%%%%%%%%%%%%%%%%%%%%%%%%%%%%%%%%%%%%
%%%%%%%%%%%%%%%%%%%%%%%%%%%%%%%%%%%%%%%%
\vspace{4mm}

\noindent{\bf Nucleation and spinodal adsorption}\\
The adsorption kinetics is classified into three distinct processes, as shown in Fig.~\ref{fig::snapshot}. When the chemical potential $\mu$ is in the metastable region (I in Fig.~\ref{fig::summary}c), nucleation-growth is observed: Adsorption domains at surfaces nucleate stochastically due to thermal fluctuations, and then the domains grow diffusively. Because the corner region can relax mechanical stress more easily than the surface region, corner nucleation occurs more frequently than surface nucleation. Thus, only the corner nucleation occurs when the chemical potential is far below the spinodal point, as shown in Fig.~\ref{fig::snapshot}a, whereas surface nucleation also occurs when approaching the spinodal point, as shown in Fig.~\ref{fig::snapshot}b (see Supplementary Videos 1 and 2). Before nucleation occurs, the adsorption fraction $n_{\rm ads}(t)$ grows as $n_{\rm ads}(t)\sim t^\gamma$ with $\gamma<1/2$, because the nucleation is a stochastic event (see Methods for the definition of $n_{\rm ads}(t)$). After the corner nucleation occurs, $\gamma$ exhibits a crossover to 1 due to the domain growth in both the $x$ and $y$ directions with $t^{1/2}$, as described in the next section (see Extended Data Fig.~\ref{fig::ex::mu_dep}).

%spinodal
Above the spinodal point (II in Fig.~\ref{fig::summary}c), on the other hand, the desorbed state becomes linearly unstable, resulting in spinodal-type domain growth. Adsorption domain growth occurs both at the corners and on the surface. However, adsorption at the surface does not proceed homogeneously, as shown in Fig.~\ref{fig::snapshot}c (see Supplementary Video 3). Coexistence of adsorbed and desorbed regions is observed. Because the desorbed sites have a smaller lattice size and are more flexible, they deform more easily than the adsorbed sites to reduce overall elastic energy, which results in the creasing of the host framework. As adsorption proceeds further, the adsorption domains merge into large clusters, and then the creasing points exhibit coarsening. In the supercritical region (III in Fig.~\ref{fig::summary}c), the host’s elasticity is no longer the dominant factor, and the adsorption process becomes diffusion under simple exclusion. In the following, we focus on the spinodal growth, whose growth law and scaling properties are governed by elastic heterogeneity.

Here, it is worth noting that the morphology of the observed patterns depends on the simulation conditions and the model parameters (see Extended Data Fig.~\ref{fig::ex::schematic}d--f and Supplementary Videos 4--6), while the boundary condition does not affect qualitative features (see Extended Data Fig.~\ref{fig::ex::snapshot_symmetric}).

%%%%%%%%%%%%%%%%%%%%%%%%%%%%%%%%%%%%%%%%
%%%%%%%%%%%%%%%%%%%%%%%%%%%%%%%%%%%%%%%%
% adsorption distribution
%%%%%%%%%%%%%%%%%%%%%%%%%%%%%%%%%%%%%%%%
%%%%%%%%%%%%%%%%%%%%%%%%%%%%%%%%%%%%%%%%

\begin{figure}[t]
\centering
\includegraphics[width=85mm]{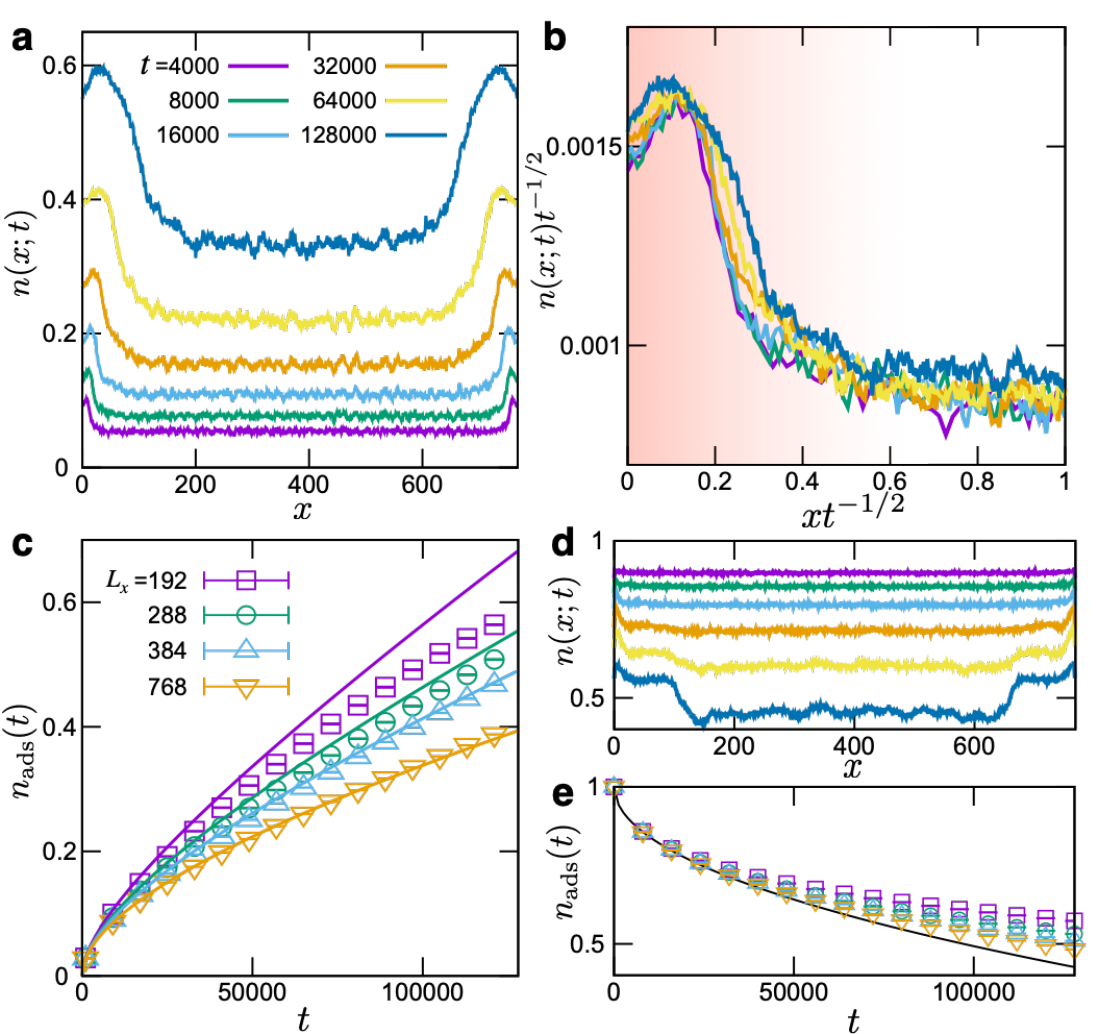}
\caption{
{\bf Adsorption distribution and adsorption fraction.}
{\bf a,} Spatial distribution of the adsorbed particle fraction with respect to $x$ in adsorption processes. Averages of 300 independent simulation runs are taken.
{\bf b,} The scaling plot $\tilde{n}(xt^{-1/2})=n(x;t)t^{-1/2}$.
{\bf c,} System size dependence of the adsorption fraction in adsorption processes. The solid curves represent Eq.(\ref{eq::adsorption}), where $A=0.000831(16)$ and $B=0.000578(54)$ are obtained by the fitting of the $L_x=768$ data.
{\bf d,} Spatial distribution of the adsorbed particle fraction in desorption processes.
{\bf e,} System size dependence of the adsorption fraction in desorption processes. The desorption decrease is slower than $t^{1/2}$, which is shown by the solid curve. The following parameters are used: $K=3.0$, $\lambda=1.4$, and $T=0.1$. Panels {\bf a,b,c} are for $\mu=1.2$ (adsorption), while panels {\bf d,e} are for $\mu=0.3$ (desorption). The system size is $L_x=768$ in panels {\bf a,b,d}.
}
\label{fig::distribution}
\end{figure}

\vspace{4mm}

\noindent{\bf Adsorption distribution}\\
For spinodal growth in Fig.~\ref{fig::snapshot}c, the scaling properties of the adsorption fraction $n_{\rm ads}(t)$ can be understood by separating the contributions of the corner and surface growths. The time evolution of adsorption distribution along the $x$ direction $n(x;t)$ is displayed in Fig.~\ref{fig::distribution}a (see Methods for the definition of $n(x;t)$). Here, averages of 300 independent runs are taken. $n(x;t)$ exhibits two peaks near the corners and a plateau in the central surface region. Although adsorption proceeds heterogeneously on the surface (see Fig.~\ref{fig::snapshot}c), the position of the creasing points (unadsorbed regions) is sample dependent, resulting in the smooth distribution of $n(x;t)$. As time proceeds, the peak position of $n(x;t)$ shifts inward, which results from the lateral growth of the corner domains.

The contribution of the corner growth on $n_{\rm ads}(t)$ is quantified by a scaling plot $\tilde{n}(xt^{-1/2})=t^{-1/2} n(x;t)$, which is shown in Fig.~\ref{fig::distribution}b. The collapse of $\tilde{n}(xt^{-1/2})$ for different $t$ implies that both the peak height and position are scaled as $t^{1/2}$. Hence, the contribution of the corner growth to $n_{\rm ads}(t)$ is proportional to $t$. $\tilde{n}(xt^{-1/2})$ increases steeply for $x t^{-1/2}\lesssim 0.4$, so the length of the corner region $x^*(t)$ is defined as $x^*(t)=0.4t^{1/2}$. Note that $x^*(t)$ does not depend on the horizontal system size $L_x$ (see Extended Data Fig.~\ref{fig::ex::distribution_size_dep}), implying that the corner contribution to $n_{\rm ads}(t)$ is proportional to $t^{1/2}\times 2x^*(t)/L_x \sim t/L_x$ for $x^*(t)\lesssim L_x/2$. For $x>x^*(t)$, data collapse in Fig.~\ref{fig::distribution}b implies $n(x;t)\sim t^{1/2}$ in the surface region. Hence, the surface contribution is proportional to $t^{1/2}\times (L_x-2x^*(t))/L_x$ for $x^*(t)\lesssim L_x/2$. Thus, the time dependence of $n_{\rm ads}(t)$ for $x^*(t) \lesssim L_x/2$ reads
\begin{equation}
n_{\rm ads}(t) = At^{1/2} + (B/L_x)t,
\label{eq::adsorption}
\end{equation}
where $A$ and $B$ are constants. Eq.(\ref{eq::adsorption}) implies that guest adsorption becomes more efficient when $L_x$ becomes smaller, which is confirmed in Fig.~\ref{fig::distribution}c. In the figure, $n_{\rm ads}(t)$ increases faster for smaller system sizes. The solid curve represents the fitting of $n_{\rm ads}(t)$ using Eq.(\ref{eq::adsorption}) for the $L_x=768$ data. The fitted curve agrees with the data for smaller $L_x$ in the early stage, whereas deviation is observed when $x^*(t)$ becomes comparable to $L_x/2$. For $x^*(t) \gtrsim L_x/2$, the domains at both corners interact with each other, resulting in the suppression of the lateral domain growth, whereas the growth in the $y$ direction remains unaffected. Accordingly, $n_{\rm ads}(t)$ follows $t^{1/2}$ at the late stage until the adsorbed domains reach the upper boundary ($y=L_y$). The time $t^*(L_x)$ at which Eq.(\ref{eq::adsorption}) deviates from $n_{\rm ads}(t)$ is of the form $t^*(L_x) = cL_x^2$, which indicates $x^*(t^*)=0.4\sqrt{c}L_x$. By setting $c=0.65$, $t^*(L_x)$ is calculated to be $24000$ and $54000$ for $L_x=192$ and $288$, respectively, showing a good agreement with those estimated from Fig.~\ref{fig::distribution}c.

The time evolution of $x^*(t)$ depends on thermodynamic parameters $(T, \mu)$. For larger $\mu$, for example, $x^*(t)/t^{1/2}$ becomes larger, implying that the lateral domain growth at the corner becomes faster as $\mu$ increases. In the limit $\mu\to\infty$, the domains at both corners merge immediately, indicating that the contribution of the corner lateral growth (the second term of Eq.(\ref{eq::adsorption})) vanishes. Thus, the domain growth law for $\mu \to \infty$ is $t^{1/2}$ (see Extended Data Fig.~\ref{fig::ex::mu_dep}). It is also worth noting that when deformation upon adsorption does not occur, i.e., $K=1$ and $\lambda=1$, lattice elasticity is irrelevant due to the absence of effective guest-guest interaction, resulting in diffusive growth $t^{1/2}$ (see Extended Data Fig.~\ref{fig::ex::uptake_homo}).

For desorption processes, the corner contribution exhibits a different tendency. Fig.~\ref{fig::distribution}d shows the spatial distribution $n(x;t)$ in a desorption process. The desorption of guest particles is slower near the corners than in the surface region. As a result, $n_{\rm ads}(t)$ decreases slower than $t^{1/2}$. This trend is enhanced for smaller system sizes, as shown in Fig.~\ref{fig::distribution}e. The opposite trend between adsorption and desorption kinetics arises from elastic heterogeneity, as explained by the Eshelby inclusion problem, a well-known concept in metallurgy~\cite{eshelby1957determination,Onukibook}. In alloys, it has been shown that elastic energy is reduced when harder domains embedded in a softer matrix have an isotropic shape, whereas softer domains in a harder matrix prefer an anisotropic, flattened shape. Since the adsorbed sites are more rigid than desorbed sites, this argument also holds in our systems~\cite{mitsumoto2023elastic,mitsumoto2024adsorption}. In the adsorption process, the adsorbed domains become isotropic, separated by unadsorbed narrow regions. Then, surface adsorption is slower than the corner adsorption when taking a statistical average. In the desorption process, on the other hand, the morphology of the desorbed domains becomes anisotropic, where desorbed sites form elongated, narrow channels, which penetrate into the bulk (see Extended Data Fig.~\ref{fig::ex::schematic}d and Supplementary Video 4). Then, surface desorption is faster than the corner desorption when taking a statistical average. Thus, elastic heterogeneity is responsible for the asymmetry in the adsorption and desorption kinetics. Furthermore, the model parameters ($K$, $\lambda$) control the role of elastic heterogeneity. When the adsorbed sites become more flexible, i.e., $K<1$, the adsorption-desorption kinetics remain asymmetric, but exhibit the opposite tendency (see Extended Data Figs.~\ref{fig::ex::schematic}e,f and~\ref{fig::ex::Fig_dist_uptake_soft}, and Supplementary Videos 5 and 6).

%%%%%%%%%%%%%%%%%%%%%%%%%%%%%%%%%%%%%%%%
%%%%%%%%%%%%%%%%%%%%%%%%%%%%%%%%%%%%%%%%
% scaling
%%%%%%%%%%%%%%%%%%%%%%%%%%%%%%%%%%%%%%%%
%%%%%%%%%%%%%%%%%%%%%%%%%%%%%%%%%%%%%%%%

\begin{table}[t]
\centering
\caption{Scaling exponents. $\alpha$: the global roughness exponent, $\beta$: the growth exponent, $z$: the dynamic exponent, $\beta'$: the anomalous growth exponent, $\alpha_{\rm loc}$: the local roughness exponent, $\alpha_{\rm s}$: the spectral exponent (see Fig.~\ref{fig::scaling} and Extended Data Figs.~\ref{fig::ex::family_vicsek_ex} and \ref{fig::ex::family_vicsek_ex2} for details).
}
\label{tab::scaling}
\begin{threeparttable}
\begin{tabular}{ccccccc}
%\begin{tabular}{|c|c|c|c|c|c|}
\hline
~ & $\alpha$ & $\beta$ & $1/z$ & $\beta'$ & $\alpha_{\rm loc}$ &  $\alpha_{\rm s}$ \\
\hline
EW\tnote{a} & 1/2 & 1/4 & 1/2 & - & - & 1/2 \\
\hline
KPZ\tnote{a} & 1/2 & 1/3 & 2/3 & - & - & 1/2 \\
\hline
linear MBE\tnote{a} & 3/2 & 3/8 & 1/4 & 1/8 & 1 & 3/2 \\
\hline
random diffusion\tnote{b} & $\frac{1}{2(1-\rho)}$ & $\frac{1}{2(2-\rho)}$ & $\frac{1-\rho}{2-\rho}$ & $\frac{\rho}{2(2-\rho)}$ & 1/2 & 1/2 \\
\hline
\hline
$T=0.08$, $\mu=1.2$ ~& 1.20 & 0.40 & 1/3 & 0.25 & 0.45 & 1.20 \\
\hline
$T=0.10$, $\mu=1.2$ ~& 1.18 & 0.39 & 1/3 & 0.25 & 0.42 & 1.18 \\
\hline
$T=0.15$, $\mu=1.3$ ~& 1.09 & 0.36 & 1/3 & 0.25 & 0.33 & 1.09 \\
\hline
$T=0.20$, $\mu=1.4$ ~& - & 0.31 & - & - & -  & - \\
\hline
$T=0.30$, $\mu=1.6$ ~& - & 0.28 & 0 & - & -  & - \\
\hline
$T=0.50$, $\mu=2.0$ ~& - & 0.27 & 0 & - & -  & - \\
\hline
$T=\infty$, $\mu=\infty$ ~& - & 1/4 & 0 & - & -  & - \\
\hline
\end{tabular}
\begin{tablenotes}
\item[a]EW, KPZ, and linear MBE stand for (1+1)-dimensional Edwards-Wilkinson, Kardar-Parisi-Zhang, and linear molecular-beam-epitaxy models, respectively~\cite{krug1991kinetic,Barabasi,lopez2005scaling}. 
\item[b]$\rho$ in exponents of the random diffusion model is the strength of the disorder given in the probability of the diffusion coefficient $P(D)\sim D^{-\rho}$ for $D<1$ and $P(D)=0$ for $D>1$~\cite{lopez1997superroughening}.
\end{tablenotes}
\end{threeparttable}
\end{table}

\vspace{4mm}

\noindent{\bf Anomalous scaling}\\
%So far, we have identified distinct adsorption kinetics between the corner ($x<x^*(t),~L_x-x^*(t)<x$) and surface ($x^*(t)<x<L_x-x^*(t)$) regions. 
We now focus on the creasing formation in the surface region, as shown in Fig.~\ref{fig::snapshot}c. This pattern is characterised by three physical quantities concerning the adsorption distribution: spatial correlation function $C(\ell;t)$, adsorption deviation $W(\ell;t)$, and the structure factor $S(k,t)$ (see Methods for their definition). The dynamic scaling analysis yields the scaling exponents characterising heterogeneous adsorption: the roughness exponent $\alpha$, the growth exponent $\beta$, the dynamic exponent $z$, the anomalous growth exponent $\beta'$, the local roughness exponent $\alpha_{\rm loc}$, and the spectral roughness exponent $\alpha_{\rm s}$. The presence of $\beta'$, $\alpha_{\rm loc}$ ($\neq \alpha$), and $\alpha_{\rm s}$ indicates the anomalous scaling (see Methods for details). Scaling exponents obtained in this study are summarised in Table~\ref{tab::scaling}, where those in other interfacial growth systems are also displayed for reference; Edwards-Wilkinson and Kardar-Parisi-Zhang models obey standard dynamic scaling, the linear molecular-beam-epitaxy model obeys superroughening, and the random diffusion model obeys intrinsic anomalous scaling~\cite{krug1991kinetic, Barabasi,lopez1997superroughening,lopez2005scaling}. Notably, our model exhibits an anomalous scaling distinct from theirs, as described below.

The spatial correlation function $C(\ell;t)$ is presented in Fig.~\ref{fig::scaling}a. There are two notable features in $C(\ell;t)$. (i) $C(\ell;t)$ has a maximum at $\ell=\ell_{\rm peak}(t)$, representing the typical distance between adsorbed domains and unadsorbed creases. As shown in Fig.~\ref{fig::scaling}b, the peak position is scaled as $\ell_{\rm peak}(t)\sim t^{1/z}$ with $1/z = 0.321(17)$, which is close to $1/3$. The growth law $t^{1/3}$ is characteristic of mass-conserved ordering dynamics~\cite{Onukibook}. This suggests that the lateral coarse-graining of adsorbed domains is governed by the internal redistribution of adsorbed particles under effective attractive interactions arising from elastic heterogeneity. Therefore, we adopt $z=3$ in the following. (ii) For sufficiently large $\ell$, spatial correlation vanishes so that $C(\ell;t)$ converges to time-dependent plateau values $C_{\rm plateau}(t)$, which represent the variance of $n(x;t)$. From Fig.~\ref{fig::scaling}b, $C_{\rm plateau}(t)\sim t^{2\beta}$ with the growth exponent $\beta=0.392(1)$. The scaling relation $\alpha=\beta z$ yields $\alpha\simeq 1.176$. The scaling function $\tilde{C}(\ell t^{-1/z})=C(\ell;t)\ell^{-2\alpha}$ reads
\begin{equation}
\tilde{C}(r) \sim
\begin{cases}
r^{-2(\alpha-\alpha_{\rm loc})} & r \ll r^* \\
r^{-2\alpha} & r \gg r^*,
\end{cases}
\label{eq::tC}
\end{equation}
where the crossover length $r^*=\ell_{\rm peak}(t)t^{-1/3}\cong 0.345$. $\tilde{C}(r)$ for $r \ll r^*$ does not exhibit a plateau, indicating $\alpha < \alpha_{\rm loc}$. $\alpha_{\rm loc}$ satisfies the scaling relation $\alpha_{\rm loc}=(\beta-\beta')z\simeq 0.426$, where $\beta'=0.250(4)$ from Fig.~\ref{fig::scaling}b. For $r\lesssim 0.1$, however, the slope slightly deviates from $-2(\alpha-\alpha_{\rm loc})$. The deviation arises from the discreteness of our model, which becomes non-negligible below five lattice spacings (see Extended Data Fig.~\ref{fig::ex::correlation_supple}). This effect reflects the fact that the morphology of small adsorption domains in the lattice model is strongly constrained by the underlying lattice structure~\cite{mitsumoto2023elastic,mitsumoto2024adsorption}.

Compelling evidence for anomalous scaling is obtained by examining the adsorption deviation $W(\ell;t)$, as shown in Fig.~\ref{fig::scaling}d--f. The scaling functions $\tilde{W}_1(t \ell^{-z})=\ell^{-\alpha}W(\ell;t)$ and $\tilde{W}_2(\ell t^{-1/z})=t^{-\beta}W(\ell;t)$ are displayed in Fig.~\ref{fig::scaling}e,f, respectively. Good data collapse is confirmed, indicating the existence of scaling functions. From the figure, the scaling functions read
\begin{align}
\tilde{W}_1(u) &\sim
\begin{cases}
u^\beta & u \ll u^* \\
u^{\beta'} & u \gg u^*,
\end{cases}
\\
\tilde{W}_2(v) &\sim
\begin{cases}
v^{\alpha_{\rm loc}} & v \ll v^* \\
\text{const} & v \gg v^*,
\end{cases}
\label{eq::tW}
\end{align}
where the crossover points $u^*=t\ell_{\rm peak}^{-z}=(r^*)^{-1/z}\cong 24.4$ and $v^*=r^*$. The growth and anomalous growth exponents are determined as $\beta=0.391(1)$ and $\beta'=0.252(3)$ from Fig.~\ref{fig::scaling}e, and the local roughness exponent is determined as $\alpha_{\rm loc}=0.424(4)$ from Fig.~\ref{fig::scaling}f, which agree with those obtained by the scaling analysis for the correlation function.

It is known that $\alpha_{\rm loc} < 1$ is a feature of intrinsic anomalous scaling (see Methods for details)~\cite{lopez1997superroughening}. However, the scaling of the structure factor $S(k;t)$ shows a different behaviour. To see this, the time evolution of the structure factor is presented in Fig.~\ref{fig::scaling}g. The structure factor has a distinct peak, yielding the correlation length $\xi=1/k_{\rm peak}$. The peak corresponds to the peak of the spatial correlation function $C(\ell;t)$. In accordance with $C(\ell;t)$, the peak of the structure factor shifts to lower wavenumbers and its height increases with time. Therefore, we regard $\ell_{\rm peak}$ as the correlation length $\xi$. From the scaling arguments, the scaling function $\tilde{S}(k t^{1/z})=k^{2\alpha+1}S(k;t)$ is given by
\begin{equation}
\tilde{S}(w) \sim 
\begin{cases}
w^{2\alpha+1} & w \ll w^* \\
w^{2(\alpha-\alpha_{\rm s})} & w \gg w^*,
\end{cases}
\label{eq::tS}
\end{equation}
where $w^*= t^{1/z}/\xi(t)=1/r^*$~\cite{ramasco2000generic}. As shown in Fig.~\ref{fig::scaling}h, $\tilde{S}(k t^{1/z})$ exhibits a plateau near $kt^{1/z}\sim1$, implying $\alpha_{\rm s}=\alpha \neq \alpha_{\rm loc}$. This feature is known for superroughening rather than intrinsic anomalous scaling, though $\alpha_{\rm loc}=1$ is required for superroughening. Thus, anomalous scaling in our model deviates from both intrinsic anomalous scaling and superroughening. It should be noted that the scaling plot $S(k;t)t^{-2\beta'}$ at high wavenumbers ($k\gtrsim 0.1$) deviates from both $-2\alpha-1$ and $-2\alpha_{\rm loc}-1$, as shown in Fig.~\ref{fig::scaling}i. As in the case of $C(\ell;t)$, the deviation results from the discreteness of our model.

The scaling exponents in Fig.~\ref{fig::scaling} depend on $(T,\mu)$, as shown in Fig.~\ref{fig::summary}c and Table~\ref{tab::scaling} (see Extended Data Figs.~\ref{fig::ex::family_vicsek_ex} and \ref{fig::ex::family_vicsek_ex2}). Furthermore, with increasing $T$ and $\mu$, the scaling exponents exhibit a crossover to supercritical values $\beta=1/4$ and $1/z=0$: no characteristic length scale exists. In this region, characteristic domain growth with elastic creasing does not occur because the adsorption chemical potential and thermal fluctuations are much greater than the elastic energy. In the crossover region between spinodal and supercritical regions, $W(\ell;t)$ cannot be scaled by any dynamic exponent, so that $1/z$ is not well-defined (see Extended Data Fig.~\ref{fig::ex::family_vicsek_ex}). It is also worth noting that the dynamical exponents depend on the simulation conditions and material parameters ($K$, $\lambda$), where the anomalous scaling does not necessarily hold (see Extended Data Fig.~\ref{fig::ex::family_vicsek_other}).

%%%%%%%%%%%%%%%%%%%%%%%%%%%%%%%%%%%%%%%%
%%%%%%%%%%%%%%%%%%%%%%%%%%%%%%%%%%%%%%%%
% Discussion, Summary
%%%%%%%%%%%%%%%%%%%%%%%%%%%%%%%%%%%%%%%%
%%%%%%%%%%%%%%%%%%%%%%%%%%%%%%%%%%%%%%%%
\vspace{4mm}

\noindent{\bf Discussion}\\
In this study, we identified two key features of molecular adsorption in soft porous crystals using a simple model incorporating elastic heterogeneity. First, surface effects enhance adsorption near corners due to local stress relaxation. This trend is consistent with experimental reports showing that surface and thin-film adsorption differ qualitatively from bulk behaviour~\cite{sakaida2016crystalline,hosono2019highly}, resulting in size-dependent uptake~\cite{sakata2013shape,krause2018effect,watanabe2024size}. Second, we observed anomalous adsorption scaling that deviates from both superroughening and intrinsic anomalous scaling. The anomalous scaling is characterised by scale separation between small and large length scales, resulting from cooperative adsorption-induced elastic creasing in the host framework. The inhomogeneous adsorption resembles simulation studies of MOFs using more realistic intermolecular potentials~\cite{rogge2019unraveling,schaper2023simulating}. Thus, MOFs with large breathing transitions, which produce significant elastic stresses, are candidates for observing the creasing instability and anomalous scaling~\cite{mitsumoto2023elastic}. To validate our model and identify relevant materials, it will be essential to examine the relationship between changes in elastic properties---elastic moduli in particular---and molecular rearrangements in host frameworks upon adsorption using experimental techniques, quantum mechanical calculations, and molecular dynamics simulations.

From a macroscopic perspective, the creasing instability observed in this study is reminiscent of pattern formation in polymeric gels~\cite{tanaka1987mechanical,Onukibook} and of mechanical instability in electrochemical storage materials such as hydrogen–metal alloys and battery electrodes~\cite{cogswell2012coherency,alefeld1978hydrogen,Onukibook}. Solvent adsorption and desorption in gels, accompanied by swelling and deswelling, also exhibit asymmetric patterns due to underlying elastic heterogeneity~\cite{tanaka1987mechanical,matsuo1992patterns,Onukibook}. However, the coarsening law in gels follows $\xi\sim t^{1/2}$ ($z=2$), and creasing sites are stiffer; in contrast, our results show $z = 3$ and indicate that creasing sites are more flexible. These differences may arise from (i) gel softening upon solvent uptake and (ii) solvent saturation, which occurs in gels but not necessarily in soft porous crystals. To systematically compare these systems, poromechanics theory under unsaturated solvent conditions offers a promising framework~\cite{Coussy2011}. Despite these differences, continuum models of adsorption in soft porous crystals, polymeric gels, and electrochemical storage materials share a common Ginzburg-Landau framework: elastic deformation of the host matrix coupled to guest uptake, incorporating spatial inhomogeneity~\cite{Onukibook}. Extending continuum theories to soft porous crystals may therefore provide deeper insights into the macroscopic and surface instabilities common to a wide class of responsive materials.

%Since molecular adsorption induces macroscopic shape transformation, it should be possible to enhance molecular adsorption by applying external mechanical stresses. An example of such phenomena is the Gorsky effect observed in hydrogen-metal systems, where macroscopic bending of substances enhances hydrogen diffusion~\cite{Onukibook,Mehrer}. Mechanical control of molecular adsorption and transport should be another future research direction. In this study, the mechanical connection between adjacent sites cannot be broken even under large deformation. That is, the perfect crystalline structure without defects is assumed. In practical substances, however, defects in crystals affect molecular adsorption and diffusion, called Cottrell atmosphere~\cite{Mehrer}. Examining the role of vacancies/interstitials, dislocations, and grain boundaries on molecular adsorption and transport should be another research direction. 

% \if0

%%%%%%%%%%%%%%%%%%%%%%%%%%%%%%%%%%%%%%%%
%%%%%%%%%%%%%%%%%%%%%%%%%%%%%%%%%%%%%%%%
% Methods
%%%%%%%%%%%%%%%%%%%%%%%%%%%%%%%%%%%%%%%%
%%%%%%%%%%%%%%%%%%%%%%%%%%%%%%%%%%%%%%%%
\clearpage
\noindent{\bf \large Methods}

\noindent{\bf Model Hamiltonian}\\
A coarse-grained square-lattice model, which incorporates adsorption-induced lattice expansion/contraction and hardening/softening, is employed in our Monte Carlo simulations~\cite{mitsumoto2023elastic}. The model consists of the translational degrees of freedom of the lattice nodes, $\bm{r}_i$ ($i=1,2,...,N_{\rm host}$), and the guest variable in each plaquette, $\sigma_{\smallsquare}$ ($\square=1,2,...,N_{\smallsquare}$) taking $1$ (adsorbed) or 0 (desorbed). $N_{\rm host}$ and $N_{\smallsquare}$ are the number of lattice nodes and plaquettes, respectively. $N_{\rm host}=(L_x+1)(L_y+1)$ and $N_{\smallsquare}=L_xL_y$. Each lattice node interacts with the nearest-neighbor (NN) and the next-nearest-neighbor (NNN) nodes. The NN and NNN elastic potentials of distance $r$ are given by $\frac{1}{2}(1-r)^2$ and $\frac{1}{2}(\sqrt{2}-r)^2$ when a guest particle is absent around the lattice nodes. Hence, the elastic potential of the plaquette $\square$ is given by $V_1(\qty{\bm{r}_{i\in\smallsquare}})=\frac{1}{4}\sum_{\rm NN}(1-r_{ij})^2+\sum_{\rm NNN}\frac{1}{2}(\sqrt{2}-r_{ij})^2$, where $\qty{\bm{r}_{i\in \smallsquare}}$ represents the positions of lattice nodes composing the plaquette $\square$. To incorporate the lattice expansion/contraction and hardening/softening, we add an elastic potential $V_2(\qty{\bm{r}_{i\in \smallsquare}})=k\qty[\frac{1}{4}\sum_{\rm NN}(1+\alpha-r_{ij})^2+\sum_{\rm NNN}\frac{1}{2}(\sqrt{2}(1+\alpha)-r_{ij})^2]$ when the guest particle is accommodated in the plaquette. Thus, the Hamiltonian is given by
\begin{align}
\mathcal{H} = \sum_{\smallsquare=1}^{N_{\smallsquare}} \Big[V_1(\qty{\bm{r}_{i\in \smallsquare}}) + \sigma_{\smallsquare} \qty[V_{2}(\qty{\bm{r}_{i\in\smallsquare}})-\mu]\Big],
\end{align}
where $\mu$ is the chemical potential for the guest particle adsorption. When the plaquette is occupied by a guest particle, the elastic potential of the plaquette reads $V_1+V_2=K\qty[\frac{1}{4}\sum_{\rm NN}(\lambda-r_{ij})^2+\sum_{\rm NNN}\frac{1}{2}(\sqrt{2}\lambda-r_{ij})^2]+\mu_c$, where $\lambda=1+k\alpha/(1+k)$ is a lattice expansion ($\lambda>1$) or contraction ($0<\lambda<1$) parameter, $K=1+k$ is a lattice hardening ($K>1$) or softening ($0<K<1$) parameter, and $\mu_c = 3K(\lambda-1)^2/(K-1)$ is the equilibrium adsorption-desorption transition point at $T=0$.

\vspace{4mm}

\noindent{\bf Ensemble}\\
For simulating a system in which the entire volume $V$ is changed by guest adsorption, an osmotic ensemble is employed~\cite{coudert2008thermodynamics,mitsumoto2023elastic}. Control parameters in this ensemble are the temperature $T$, the chemical potential of guest particles $\mu$, the number of nodes $N_{\rm host}$, and external pressure $P$. The osmotic grand potential is defined as $\Omega = U-TS+PV-\mu N_{\rm ads}$, where $U$ is the internal energy, $S$ is the entropy, and $N_{\rm ads}$ is the number of adsorbed particles. In the differential form, $d\Omega=-SdT+VdP-N_{\rm ads}d\mu+\mu_{\rm host}dN_{\rm host}$, where $\mu_{\rm host}$ is the chemical potential of the host framework. We fix $P=0$ for our convenience. In this study, particle adsorption proceeds from the bottom boundary; the density of guest particles becomes heterogeneous during the adsorption kinetics. Then, the bottom boundary is more swelled (contracted) than the top boundary when $\lambda>1$ ($\lambda<1$). To avoid the numerical difficulties arising from the lattice mismatch between the bottom and top boundaries, the open boundary condition, rather than the periodic boundary condition, is adopted.

\vspace{4mm}

\noindent{\bf Phase diagram}\\
To obtain the phase diagram in Fig.~\ref{fig::ex::schematic}c, we adopt a periodic boundary condition to reduce the surface effect, which alters the nature of the transition. When the periodic boundary condition is imposed, an additional degree of freedom, average swelling ratio $a$, is necessary to incorporate the dynamical change of the entire volume. Then the system volume is represented as $V=aL_xL_y$. Unit Monte Carlo step (MCS) consists of one Metropolis sweep for guest particles $\{\sigma_{\smallsquare}\}$ and $L$-times Metropolis sweeps for lattice sites $\{\bm{r}_i\}$, and $L$-times Metropolis updates for the average swelling ratio $a$. The updates of $\bm{r}_i$ and $a$ are restricted to $|\Delta \bm{r}_i| < 0.1$ and $\Delta a < 0.01$, respectively. The system size $L_x=L_y=24$ is adopted.

The spinodal curves are obtained by standard Monte Carlo simulations for $\mu$-increasing/decreasing with a change in chemical potential $\Delta \mu=0.01$ every $2\times10^4$ MCSs. Then, the hysteretic adsorption and desorption transitions are observed. Because metastable states are robust against thermal fluctuations, we regard each transition point as the spinodal point~\cite{mitsumoto2023elastic}.

Multicanonical Monte Carlo simulation is utilised to obtain the thermodynamic equilibrium phase boundary. Here, the Wang-Landau method is applied, which efficiently samples the density of energy states $g(E)$~\cite{wang2001efficient,Landau-Binder}. By adopting the weight proportional to $e^{-g(E)}$, all the energy states, including those which are rarely realized in canonical ensembles, are uniformly sampled.
% The procedure of the Wang-Landau method is as follows. {\bf i}) Set the histogram $H(E)=0$, the density of states $g(E)=0$, and parameter $\Delta g=1$. {\bf ii}) Update the microscopic state $i$ with metropolis rule $P(i\to j)=\text{min}[1,e^{g(E_i)-g(E_j)}]$, where $j$ denotes a trial state. Then, update the histograms $H(E)$ and $g(E)$ of the new state, which are increased by one and $\Delta g$, respectively. {\bf iii}) Continue {\bf ii} until $H(E)$ is sufficiently flat. {\bf iv}) Reset the histogram $H(E)=0$ and $\Delta g$ is divided by 2. {\bf v}) Repeat {\bf ii-iv} until $\Delta g$ is smaller than $10^{-6}$.
We obtain the energy histogram $H(E)$ by multicanonical Monte Carlo simulation. Then, the equilibrium probability distribution of the energy at each temperature $T$ is computed as $P(E) = H(E)e^{-E/T+g(E)}/\sum_E H(E)e^{-E/T+g(E)}$. The equilibrium phase transition point is determined as the temperature at which an equally weighted double-peak probability distribution is realised.

\vspace{4mm}

\noindent{\bf Dynamic Monte Carlo simulations}\\
We perform dynamic Monte Carlo simulations to study adsorption-desorption kinetics. The unit MCS during adsorption simulations consists of $N_{\rm host}\times L_x$ updates for lattice site positions and $N_{\smallsquare}$ updates for guest variables. In the former, a lattice node position $\bm{r}_i$ and its trial position $\bm{r}_i'=\bm{r}_i+\Delta\bm{r}$ are randomly chosen, where $|\Delta\bm{r}|<0.1$ and bond intersection is rejected to preserve the square lattice configuration. The acceptance is determined by the Metropolis rule. In the latter, a plaquette $\square$ and its neighbour $\square'$ are randomly selected. The Kawasaki dynamics is adopted~\cite{Landau-Binder}, where a trial exchange of $\square$ and $\square'$ is evaluated under the Metropolis rule with the energy change $(\sigma_{\smallsquare'}-\sigma_{\smallsquare})(V_{2}(\{\bm{r}_{i \in \smallsquare}\})-V_{2}(\{\bm{r}_{i\in\smallsquare'}\}))$, which is not affected by the chemical potential $\mu$. If the plaquette $\square$ is on the bottom boundary, a trial step also includes particle entrance/exit through the boundary. If particle entrance/exit is selected, $\sigma_{\smallsquare}$ is updated using the Metropolis rule with an energy change $\Delta E = (2\sigma_{\smallsquare}-1)[V_{2}(\{\bm{r}_{i \in \smallsquare}\})-\mu]$, otherwise, the Kawasaki dynamics is adopted in the same manner as above. Before performing adsorption simulations, 48,000 single lattice-site updates are conducted to realise initial equilibrium states. The vertical system length is fixed to $L_y=48$. For each parameter, 300 independent simulation runs are conducted to obtain their sample averages.

\vspace{4mm}

\noindent{\bf Adsorption distribution and correlation functions}\\
The adsorption fraction and distribution are defined by $n_{\rm ads}(t)=(1/N_{\smallsquare})\sum_{\smallsquare}\sigma_{\smallsquare}(t)$ and $n(x;t)=(1/L_y)\sum_y\sigma_{\smallsquare}(t)$, respectively. To quantify the inhomogeneity of $n(x;t)$, the spatial correlation function $C(\ell;t)$, the adsorption deviation $W(\ell;t)$, and the structure factor $S(k,t)$ are calculated. They are defined by
\begin{align}
C(\ell;t)&=\frac{1}{\ell_{\rm surf}(t)+1-\ell}\sum_x [n(x;t) - n(x+\ell;t)]^2,\label{eq::C}\\
W(\ell;t)&=\frac{1}{\ell_{\rm surf}(t)-\ell}\sum_x \sqrt{\expval{[n(x;t)-\expval{n(x;t)}_\ell]^2}_\ell},\label{eq::W}\\
S(k;t)&=\qty|\tilde{n}(k;t)|^2,\label{eq::S}
\end{align}
where $\sum_{x}$ is the summation over the surface region $x^*(t)<x<L_x-x^*(t)-\ell$, $\ell_{\rm surf}(t)=L_x-2x^*(t)$ is the surface length unaffected by corner adsorption, $\expval{n(x;t)}_\ell=\sum_{x'=0}^\ell n(x+x';t) / (\ell+1)$ denotes a segmental average, and $\tilde{n}(k;t)=\sum_{x=x^*(t)}^{L_x-x^*(t)}\Delta n(x;t)e^{2\pi i k x} / \sqrt{\ell_{\rm surf}+1}$ is the Fourier transform of the adsorption distribution with $\Delta n(x;t)=n(x;t)-\sum_{x=x^*(t)}^{L_x-x^*(t)}n(x;t) / (\ell_{\rm surf}(t)+1)$.

\vspace{4mm}

\noindent{\bf Dynamic scaling}\\
Interfacial growth systems often exhibit self-affinity, meaning that interfacial fluctuations are statistically invariant under anisotropic scale transformations~\cite{krug1991kinetic,Barabasi}. More precisely, under a lateral scale transformation $x\to bx$, the interfacial fluctuation $\Delta n(x;t)$ is rescaled as $\Delta n\to b^{\alpha}\Delta n$. $\alpha$ is called the roughness exponent, restricted to $\alpha<1$~\cite{amar1993groove}. Since the time evolution must remain invariant under this transformation, the time $t$ is rescaled as $t\to b^z t$, where $z$ is called the dynamic exponent. Accordingly, under a temporal scale transformation $t\to b't$, the lateral length scale and the interfacial fluctuation are rescaled $x\to b'^{1/z}x$ and $\Delta n\to b'^{\beta}\Delta n$, respectively, where $\beta=\alpha/z$ is called the growth exponent. Thus, the scaling form of $\Delta n$ reads $\Delta n(x;t)=b^{-\alpha}\Delta n(bx;b^zt)=b'^{-\beta}\Delta n(b'^{1/z}x;b't)$. In experiments and numerical simulations on interfacial growth, these exponents are extracted from the dynamic scaling of $C(\ell;t)$, $W(\ell;t)$, and $S(k,t)$, which are defined in Eqs. (\ref{eq::C})--(\ref{eq::S}). The scaling forms are given by
\begin{align}
C(\ell;t)&=\ell^{2\alpha}\tilde{C}(\ell t^{-1/z}),\\
W(\ell;t)&=\ell^\alpha \tilde{W}_1(t \ell^{-z})=t^\beta \tilde{W}_2(\ell t^{-1/z}),\\
S(k;t)&=k^{-(2\alpha+1)}\tilde{S}(k t^{1/z}).
\end{align}
All scaling functions, $\tilde{C}$, $\tilde{W}_1$, $\tilde{W}_2$, and $\tilde{S}$, are also characterised by $\alpha,\beta$ and $z$. Typical examples that exhibit self-affinity are the Edwards-Wilkinson~\cite{edwards1982surface} and Kardar-Parisi-Zhang (KPZ) models~\cite{kardar1986dynamic}, whose scaling exponents are listed in Table~\ref{tab::scaling}.

However, self-affinity does not necessarily hold~\cite{lopez1997superroughening,ramasco2000generic}. For example, an experiment on fluid imbibition into a porous medium reports anomalous scaling, in which local (short length scale) and global (large length scale) interfacial fluctuations are characterised by different roughness exponents~ \cite{soriano2005anomalous}. In such cases, the global roughness exponent $\alpha$ is not restricted to $\alpha<1$. Anomalous scaling is characterised by three additional scaling exponents~\cite{ramasco2000generic}: local roughness exponent $\alpha_{\rm loc}<\alpha$, anomalous growth exponent $\beta'>0$, and spectral roughness exponent $\alpha_{\rm s}$, satisfying the scaling relation $\alpha_{\rm loc}=z(\beta-\beta')$. $\alpha_{\rm loc}$ and $\beta'$ appear in $\tilde{C}$ and $\tilde{W}$, whereas $\alpha_{\rm s}$ appears in $\tilde{S}$, as shown in Eqs.~(\ref{eq::tC})--(\ref{eq::tS}). Standard dynamic scaling is recovered when $\alpha_{\rm loc}=\alpha$, $\beta'=0$ and $\alpha_{\rm s}=\alpha$. The anomalous scaling has been classified into superroughening and intrinsic anomalous scaling~\cite{lopez1997superroughening}. Superroughening is characterised by $\alpha_{\rm loc}= 1$ and $\alpha>1$. The structure factor exhibits the ordinary scaling, i.e., $\alpha_{\rm s} = \alpha$. It is realised for a system possessing conservation laws~\cite{lopez2005scaling}, such as the linear molecular beam epitaxy model~\cite{herring1951, mullins1957theory, wolf1990growth, dassarma1991new}. In the case of intrinsic anomalous scaling, the local exponent $\alpha_{\rm loc}$ is directly linked to the spectral roughness exponent $\alpha_{\rm s}$, i.e., $\alpha_{\rm loc}=\alpha_{\rm s}$. The local exponent takes $\alpha_{\rm loc}<1$ and $\alpha$ can take any value. It is realised for systems having nonlocality or quenched disorder~\cite{lopez2005scaling}, e.g., the random diffusion model~\cite{lopez1997superroughening}, the fractional KPZ model with long-range spatially correlated noise~\cite{xia2016numerical}, and a fluid imbibition experiment~\cite{soriano2005anomalous}. 

However, as shown in the main text, $\alpha_{\rm loc}<1<\alpha$ and $\alpha_{\rm s}=\alpha$ are realised in our model. Thus, adsorption kinetics in our model exhibits anomalous scaling distinct from both superroughening and intrinsic anomalous scaling.

\vspace{4mm}

\noindent{\bf Landau theory of adsorption transition coupled to elasticity}\\
To understand the adsorption-desorption transition with hysteresis in Fig.~\ref{fig::summary}c, it is convenient to construct a Landau free energy of the adsorption fraction $\phi$ ($n_{\rm ads}$ in the main text) coupled with elastic deformation. For this purpose, spatial inhomogeneity is not incorporated. Using the lattice displacement vector $\bm{u}$, the elastic strains in two dimensions read $e_1=\nabla\cdot \bm u$, $e_2=\nabla_x u_x-\nabla_y u_y$, and $e_4=\nabla_x u_y + \nabla_y u_x$. $e_1$ is the volumetric strain, and $e_2$ and $e_4$ are the shear strains. In the absence of direct interactions between adsorbates, the Landau free energy reads 
\begin{equation}
f=f_0(\phi) - \mu_{\rm ads} \phi -\alpha \phi e_1 +f_{\rm el}(\phi,e),
\end{equation}
where $f_0(\phi)=k_{\rm B}T[\phi\ln \phi + (1-\phi)\ln (1-\phi)]$ is the translational entropy of the adsorbates, $\mu_{\rm ads}$ is the adsorption chemical potential ($\mu$ in the main text), and $\alpha$ represents the coupling between lattice deformation and molecular adsorption. $\alpha>0$ for the expansion case, and $\alpha<0$ for the contraction case. $f_{\rm el}(\phi,e)=\frac{K(\phi)}{2}e_1^2 + \frac{G(\phi)}{2}(e_2^2+e_4^2)$ is the isotropic elastic energy, where $K$ is the bulk modulus and $G$ is the shear modulus. By incorporating adsorption-induced lattice hardening/softening, the elastic coefficients $K$ and $G$ depend on $\phi$ as $K=K_0+K_1\phi$ and $G=G_0+G_1\phi$. Under the mechanical equilibrium condition with stress-free boundary conditions, the space average of the free energy variation with respect to the elastic tensor vanishes, which yields $\av{Ge_2}=\av{Ge_4}=0$, and $\av{e_1}=\alpha\av{\phi/K}$.

When the system is at chemical and mechanical equilibrium, $\phi$ and $e$ become homogeneous; hence, the Landau free energy reads
\begin{equation}
f=k_{\rm B}T[\phi\ln \phi + (1-\phi)\ln (1-\phi)] - \mu_{\rm ads}\phi - \frac{\alpha^2}{2K}\phi^2,
\end{equation}
by eliminating the elastic field. The last term describes the attractive guest-guest interaction due to the presence of $\alpha$. The chemical equilibrium is given by $d f/d \phi=0$, which reads
\begin{equation}
k_{\rm B}T\ln \frac{\phi}{1-\phi} -\frac{\alpha^2}{K}\phi + \frac{\alpha^2 K_1}{2K^2}\phi^2 =\mu_{\rm ads}.
\label{eq::ChemEquiv}
\end{equation}
We consider three cases: (i) $\alpha=0$, that is, lattice expansion/contraction does not occur. The left-hand side increases monotonically with $\phi$, yielding $\phi=1/(1+e^{-\beta\mu_{\rm ads}})$ without a phase transition (see Extended Data Fig.~\ref{fig::ex::uptake_homo}). (ii) $\alpha\neq 0$ and $K_1=0$. Due to the second term, the l.h.s. does not increase monotonically when $\alpha^2\ge 4k_{\rm B}TK$, resulting in the appearance of a phase transition regardless of the sign of $\alpha$. (iii) $\alpha\neq 0$ and $K_1\neq 0$. The sign of the third term depends on the sign of $K_1$. The elasticity-mediated attractive interaction is enhanced for $K_1<0$, whereas it is suppressed for $K_1>0$; thus, when adsorbed sites are more flexible than desorbed sites, they form a spatially connected cluster (see Extended Data Fig.~\ref{fig::ex::schematic}e), whereas adsorbed domains are separated by unadsorbed regions when adsorbed sites are more rigid (see Fig.~\ref{fig::snapshot}c).

%In the kinetic process, we need to solve kinetic equations regarding $\phi$ and $\bm u$. The former is described by the Cahn-Hilliard equation with surface chemical reaction as
%\begin{equation}
%\frac{\partial}{\partial t}\phi = \nabla\cdot (L\nabla\frac{\delta {\cal F}}{\delta \phi} - {\bm J}^{\rm R}) +\delta(y)(-L_0\frac{\delta {\cal F}}{\delta \phi} + f^{\rm R}),
%\label{eq::phi}
%\end{equation}
%where we assume constant kinetic coefficient $L$ and $L_0$, and ${\bm J}^{\rm R}$ and $f^{\rm R}$ are the thermal noises satisfying $\langle J^{\rm R}_i({\bf r},t)J^{\rm R}_j({\bf r'},t)\rangle = 2Lk_{\rm B}T\delta_{ij}\delta({\bf r}-{\bf r'})\delta(t-t')$, $\langle f^{\rm R}(x,t)f^{\rm R}(x',t)\rangle = 2L_0k_{\rm B}T\delta(x-x')\delta(t-t')$. The elastic equations read
%\begin{equation}
%\frac{\partial^2}{\partial t^2}\bm u =\nabla\cdot \sigma +\eta\nabla^2 \bm v.
%\label{eq::u}
%\end{equation}
%where $\sigma_{\mu\nu}=\delta {\cal F}/\delta\nabla_\mu u_\nu$ is the stress tensor, and $\eta$ represents the solid viscosity~\cite{Landau7}, which is necessary to stabilise numerical simulations. Here, we neglect the random stress tensor arising from $\eta$ because we are interested in mechanically equilibrium conditions.

It should be noted that, in general, the thermodynamic phase diagram does not necessarily correspond to the instability upon molecular adsorption through the surface, because the shear deformation near the surface is also important~\cite{onuki1993theory}. In the present study, however, the shear modulus and bulk modulus show the same dependence on adsorption fraction. Then the difference in macroscopic instability and surface instability points is negligible.

Thus, the Landau free energy without spatial inhomogeneity can describe the thermodynamic adsorption–desorption transition. However, it does not account for the adsorption kinetics we have investigated in this study. A detailed analysis of the Ginzburg–Landau free energy, incorporating spatial inhomogeneity, would offer deeper insights into the physical origin of the anomalous dynamic scaling discussed in the main text.

\vspace{2mm} 
\noindent{\bf Data availability}\\
% All data are included in the article and supporting information.
Input files to generate all of the figures are openly available at GitHub (https://github.com/kmitsumoto51/mof\_kinetics).

\vspace{2mm} 
\noindent{\bf Code availability}\\
The computer codes used in this study are available from the corresponding author upon request.

\vspace{4mm}
\noindent 
{\bf Acknowledgments}\\
The authors would like to thank Kazumasa A. Takeuchi and Tetsuo Yamaguchi for valuable discussions.
This work was supported by the JSPS KAKENHI Grant No. JP24K00594, JP25H01978, and JP25K17354.

\vspace{2mm} 
\noindent 
{\bf Author contributions:} K.M. and K.T. conceived the project, K.M. performed numerical simulations and analysed the data, and K.M. and K.T. discussed the results and wrote the manuscript.

\vspace{2mm} 
\noindent 
{\bf Competing interests.}\\
The authors declare no competing financial interests.

\vspace{2mm} 
\noindent 
{\bf Corresponding author.}
Correspondence and requests for materials should be addressed to K.M. and K.T. 

\bibliographystyle{naturemag_noURL}
%\bibliography{ref-PCPMOF2}% Produces the bibliography via BibTeX.

\clearpage

\renewcommand{\figurename}{{\bf Extended Data Fig.~}}
\setcounter{figure}{0}

%1
\begin{figure*}[t]
\centering
\includegraphics[width=170mm]{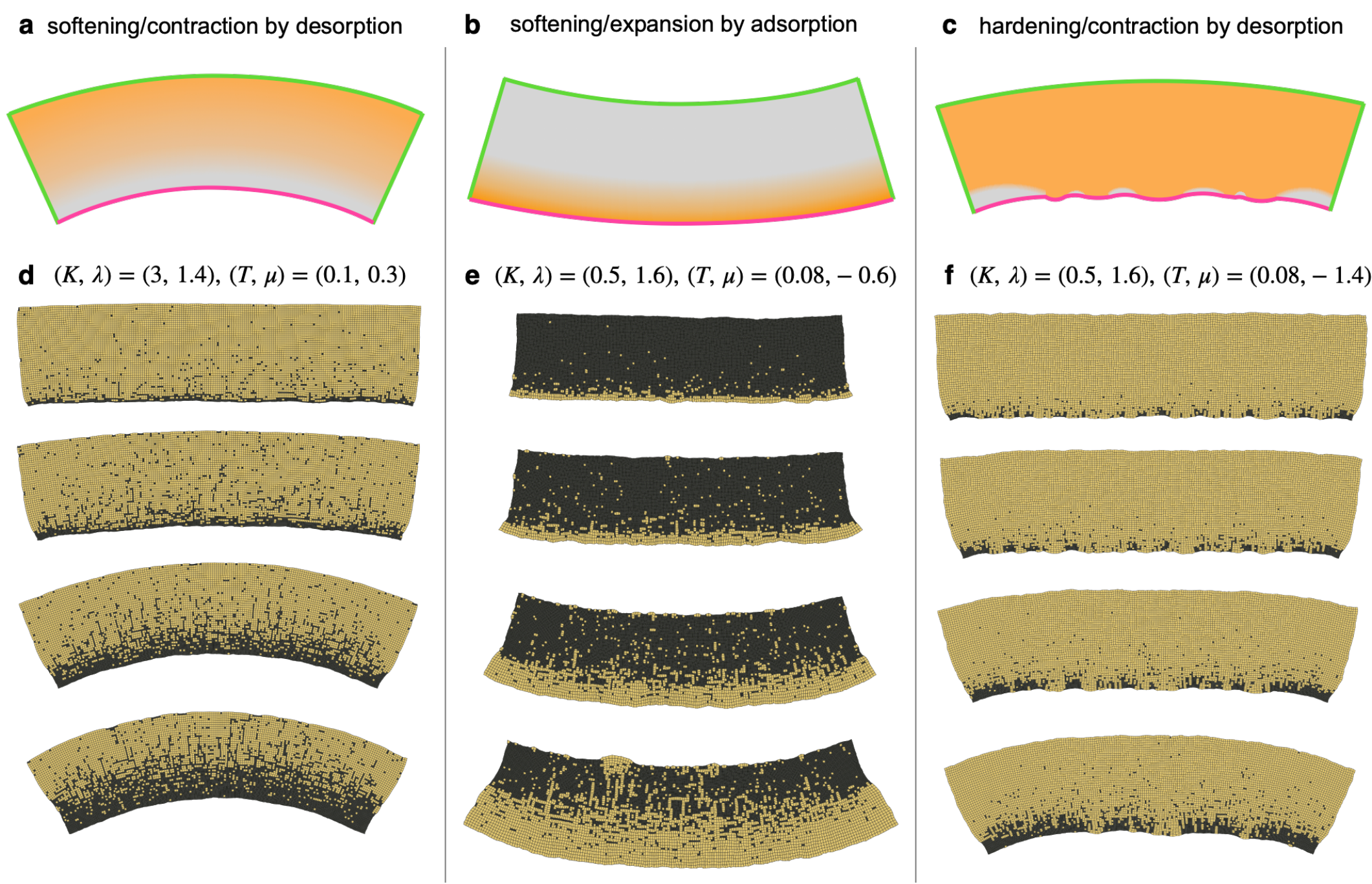}
\caption{
{\bf Schematic pictures of surface patterns and snapshots for various parameters.}
{\bf a-c,} Schematic pictures of pattern formation. {\bf a,} Desorption process for a system with lattice hardening and expansion upon adsorption ($K>1, \lambda>1$). {\bf b,} Adsorption process for a system with lattice softening and expansion upon adsorption ($0<K<1, \lambda>1$). {\bf c,} Desorption process for the same system as in {\bf b}. In these schematics, adsorbed (desorbed) regions are displayed in orange (gray). Guest particles enter and exit from the bottom boundary, indicated by the magenta line. The other boundaries, represented by green lines, do not permit the entrance or exit of particles, while stress relaxation occurs.
{\bf d,} Snapshots of a desorption process corresponding to {\bf a} for $K=3.0$ and $\lambda=1.4$. $(T,\mu)=(0.1,0.3)$, where desorption proceeds in a spinodal manner (see Fig.~\ref{fig::ex::schematic}c). $t=4000, 16000, 64000, 256000$ from the top to the bottom (see Supplementary Video 4). Since the desorbed sites are more flexible than the adsorbed sites, the growing domain favours an elongated shape along the surface to reduce the elastic stress. The elongation of the softer domain results from elastic heterogeneity. The domain becomes uniform in the lateral dimension, resulting in the absence of creases. Although the chemical potential is far below the spinodal point in the figure, qualitatively the same kinetics is confirmed when approaching the spinodal point.
{\bf e,} Snapshots of an adsorption process corresponding to {\bf b} for $K=0.5$ and $\lambda=1.6$, indicating that adsorbed sites expand and become more flexible. $(T,\mu)=(0.08,-0.6)$ to examine adsorption kinetics. $t=4000, 16000, 64000, 256000$ from the top to the bottom (see Supplementary Video 5). Since the adsorbed sites are more flexible, the same argument as {\bf d} holds, leading to the laterally uniform distribution of the adsorbed domain.
{\bf f,} Snapshots of a desorption process corresponding to {\bf c} for $K=0.5$ and $\lambda=1.6$. $(T,\mu)=(0.08,-1.4)$  to examine desorption kinetics. $t=8000, 16000, 32000, 64000$ from the top to the bottom (see Supplementary Video 6). Contrary to {\bf d} and {\bf e}, the growing desorbed sites are more rigid. As a manifestation of elastic heterogeneity, harder domains favour droplet-like compact shapes. Then, similar to Fig.~\ref{fig::snapshot}c in the main text, the desorption distribution becomes heterogeneous in the lateral dimension, resulting in anomalous scaling (see Extended Data Fig.~\ref{fig::ex::family_vicsek_other}). Because the adsorbed, flexible sites have a larger volume than the desorbed sites, an undulating structure rather than creases forms. The system size for all snapshots is $L_x=192$.
}
\label{fig::ex::schematic}
\end{figure*}

%2
\begin{figure*}[t]
\centering
\includegraphics[width=160mm]{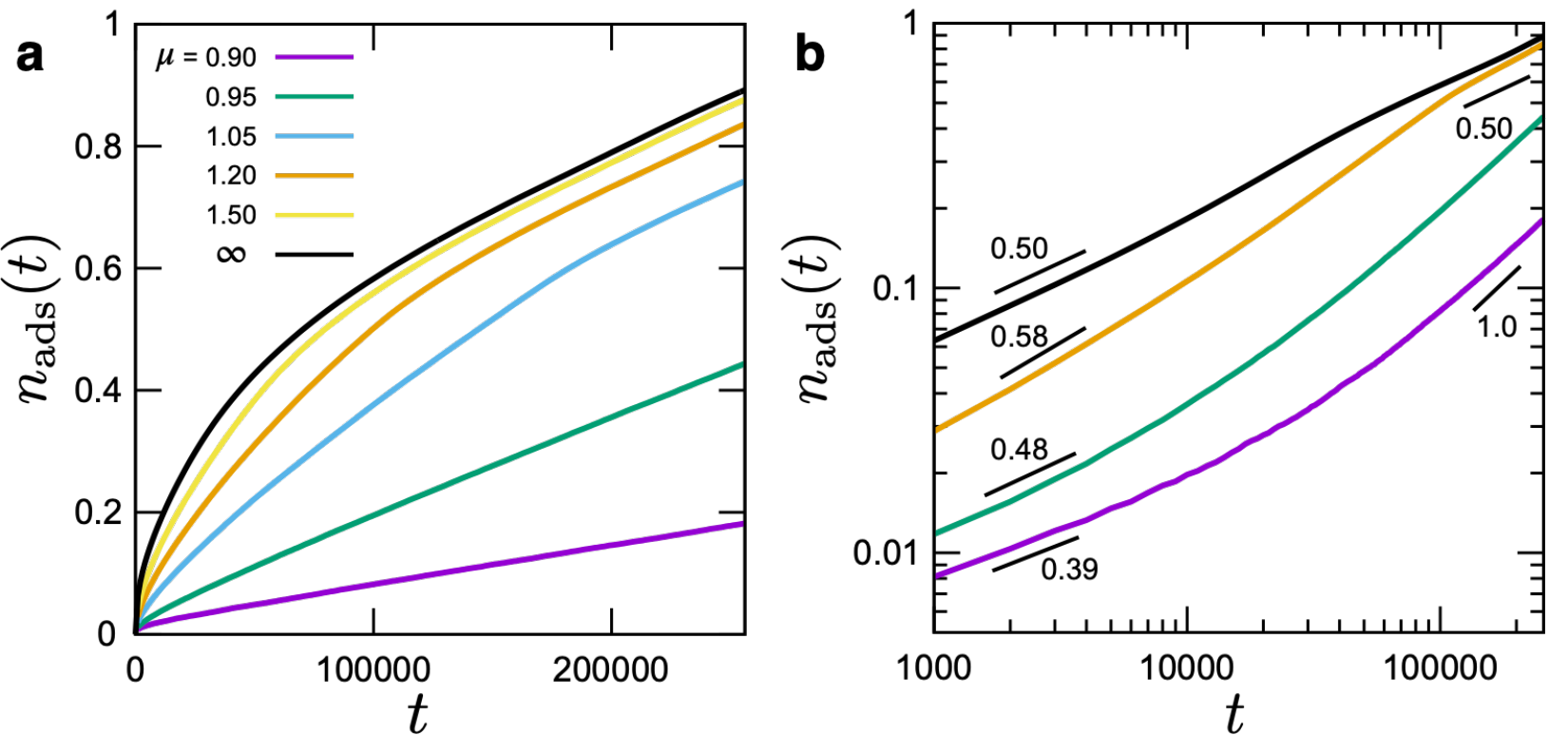}
\caption{
{\bf Chemical potential dependence of the adsorption kinetics.}
{\bf a,b,} The time evolution ({\bf a}) and its log-log plot ({\bf b}) of the adsorption fraction $n_{\rm ads}(t)$ for various $\mu$ at $T=0.1$ for $K=3$ and $\lambda=1.4$. When $\mu$ is below the spinodal $\mu_{\rm s}=0.98$, nucleation at the corners and surfaces occurs stochastically, resulting in slower growth. After the nucleation, the corner domains grow laterally and vertically with $t^{1/2}$, resulting in $n_{\rm ads}\sim t$ until the domains reach the top boundary. When $\mu$ is above the spinodal, Eq.(\ref{eq::adsorption}) in the main text holds for $t$ smaller than $t^*$; then $n_{\rm ads}(t)$ exhibits a crossover to the diffusive growth $t^{1/2}$. When $\mu$ is sufficiently large ($\mu=\infty$: black curve), $n_{\rm ads}(t)$ grows uniformly in the $x$ direction, resulting in diffusive growth $t^{1/2}$ until adsorbates reach the top boundary. The system size is $L_x=192$.
}
\label{fig::ex::mu_dep}
\end{figure*}

%3
\begin{figure*}[t]
\centering
\includegraphics[width=140mm]{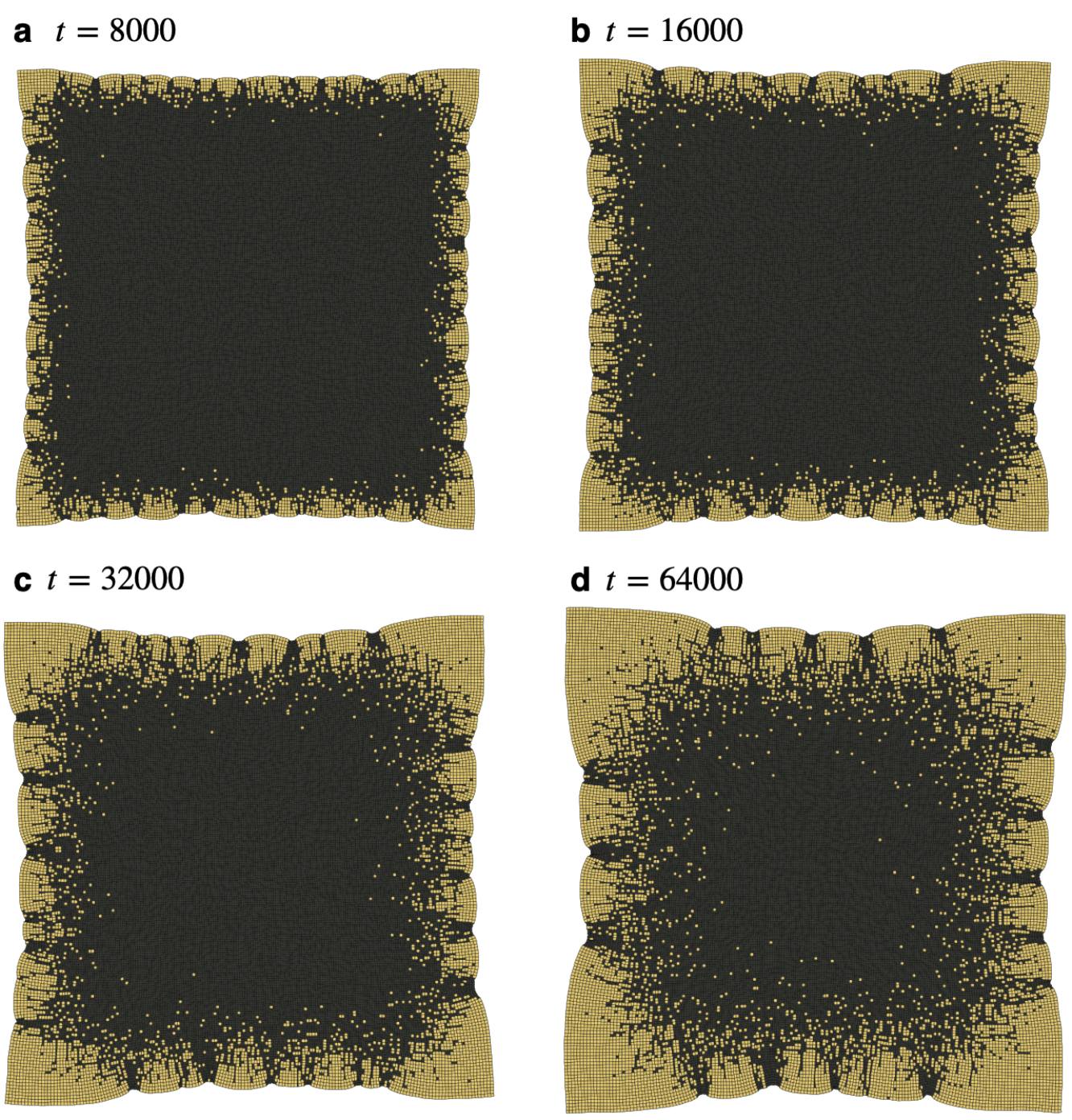}
\caption{
{\bf Adsorption kinetics using a different boundary condition.} 
{\bf a-d,} Snapshots of an adsorption process at $t=8000$ ({\bf a}), $t=16000$ ({\bf b}), $t=32000$ ({\bf c}), $t=64000$ ({\bf d}). In this simulation, guest particles enter from all boundaries. Similar to Fig.~\ref{fig::snapshot}c in the main text, large domains at the corners, adsorbed islands on the surfaces, and unadsorbed narrow regions exhibiting elastic creases are observed. The simulation was performed in the spinodal growth region with parameters $(T,\mu)=(0.1,1.2)$, $K=3.0,~\lambda=1.4$, and a system size of $L_x=L_y=192$.
}
\label{fig::ex::snapshot_symmetric}
\end{figure*}

%4
\begin{figure*}[t]
\centering
\includegraphics[width=160mm]{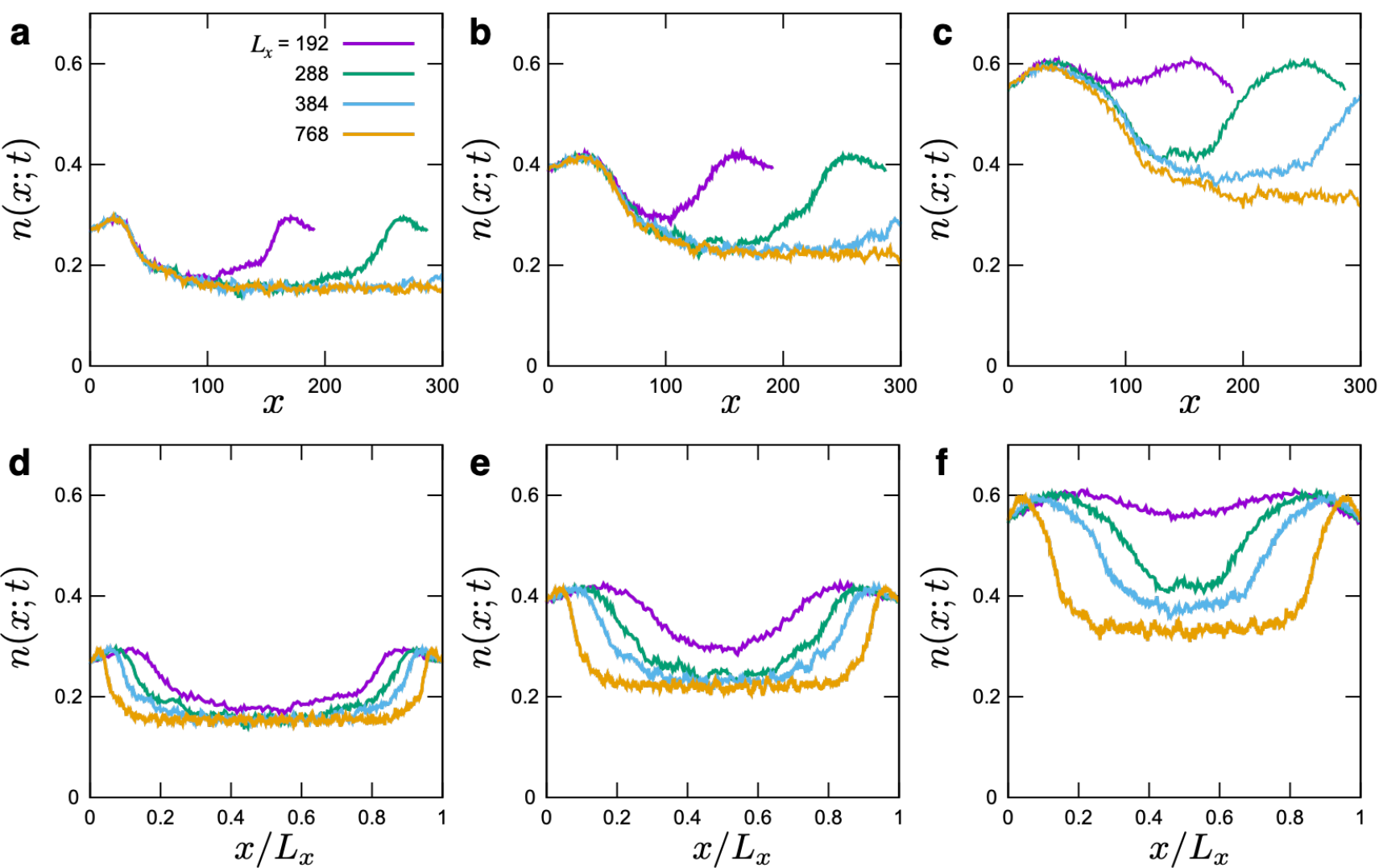}
\caption{
{\bf System size dependence of the adsorption distribution.}
{\bf a-c,} Distribution function of adsorbed particles $n(x;t)$ along the $x$ direction at $t = 32000$ ({\bf a}), 64000 ({\bf b}), and 128000 ({\bf c}) for $L_x = 192, 288, 384, 768$. $n(x;t)$ has the same value close to the left corner ($x=0$), indicating that the length of the corner region $x^*(t)$ is independent of the lateral system length $L_x$. 
{\bf d-f,} $n(x;t)$ with respect to the normalised coordinate $x/L_x$ at $t = 32000$ ({\bf d}), 64000 ({\bf e}), and 128000 ({\bf f}). Since $x^*(t)$ does not depend on $L_x$, the size of corner domains is independent of the system size a given time; hence, the corner contribution to the adsorption fraction $n_{\rm ads}(t)$ becomes smaller as $L_x$ increases, as shown in Fig.~\ref{fig::distribution}c. The other simulation parameters are the same as those in Fig.~\ref{fig::distribution}a in the main text.
}
\label{fig::ex::distribution_size_dep}
\end{figure*}

%5
\begin{figure*}[t]
\centering
\includegraphics[width=170mm]{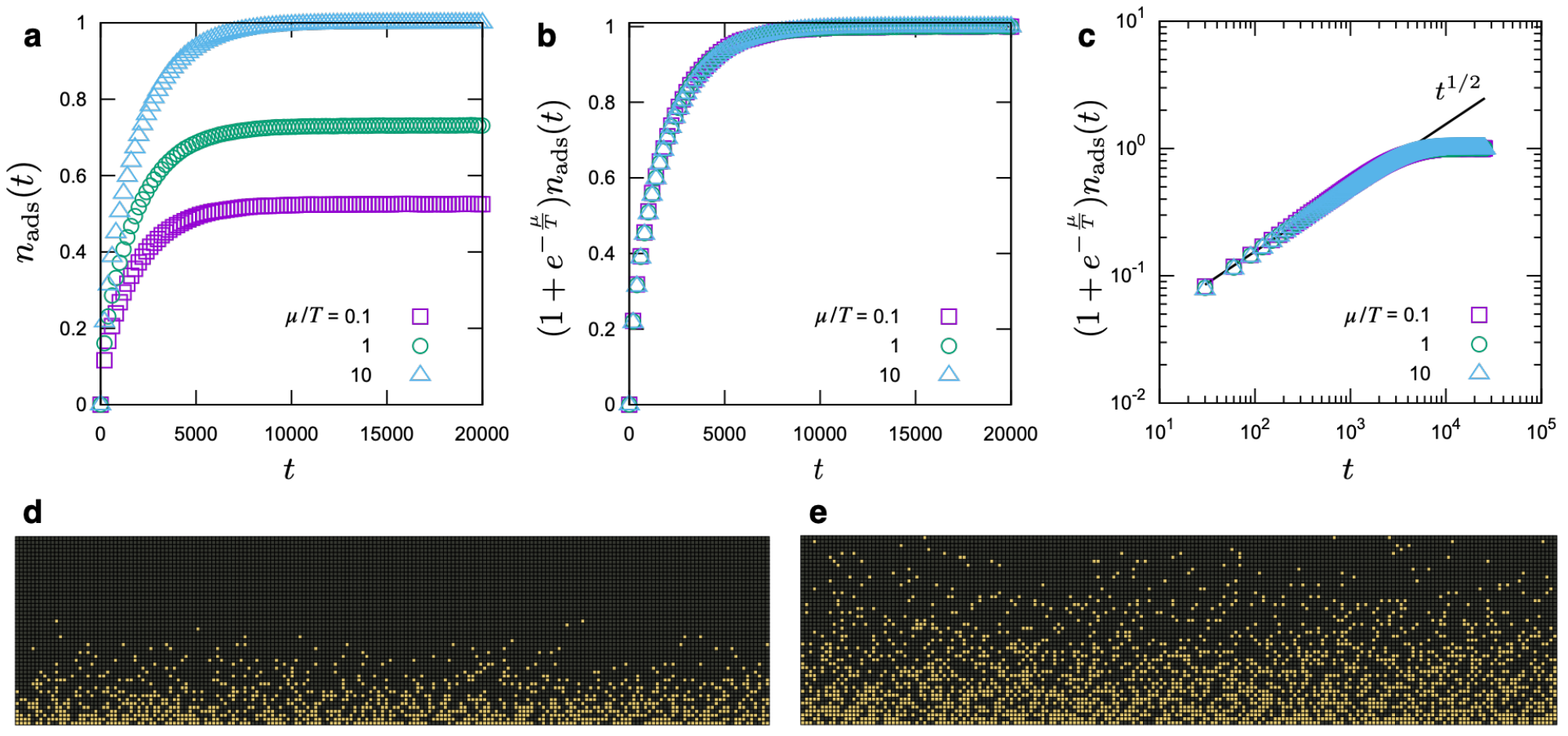}
\caption{
{\bf Adsorption fraction and snapshots without host's elastic interaction.}
{\bf a,} Time evolution of the adsorption fraction for $K=\lambda=1$, which indicates the absence of lattice deformation and rigidity changes upon adsorption. There is no elasticity-mediated interaction between guest particles (see also Methods), corresponding to the rigid limit of the host framework. The kinetic process becomes identical to the simple exclusion processes with a boundary in contact with a particle bath. Guest particles move diffusively inside so that $n_{\rm ads}(t)$ is proportional to $t^{1/2}$ uniformly in the $x$ direction, and converges to the Langmuir isothermal equilibrium value $1/(1+e^{-\mu/k_{\rm B}T})$.
{\bf b,c,} The scaling plot using $1/(1+e^{-\mu/k_{\rm B}T})$ ({\bf b}) and its log-log plot ({\bf c}). The scaling plot exhibits a good collapse. The adsorption fraction increases as the square root of time $t$ before reaching a saturation point. 
{\bf d,e,} Snapshots of the adsorption process for $\mu/T=1$ at $t=100$ ({\bf d}) and $t=400$ ({\bf e}), where particle migration occurs uniformly without forming domains. Thus, both the adsorption growth law and pattern formation described in the main text are absent when the host framework is sufficiently rigid without breathing upon molecular uptake. The system size is $L_x=192$.
}
\label{fig::ex::uptake_homo}
\end{figure*}

%6
\begin{figure*}[t]
\centering
\includegraphics[width=120mm]{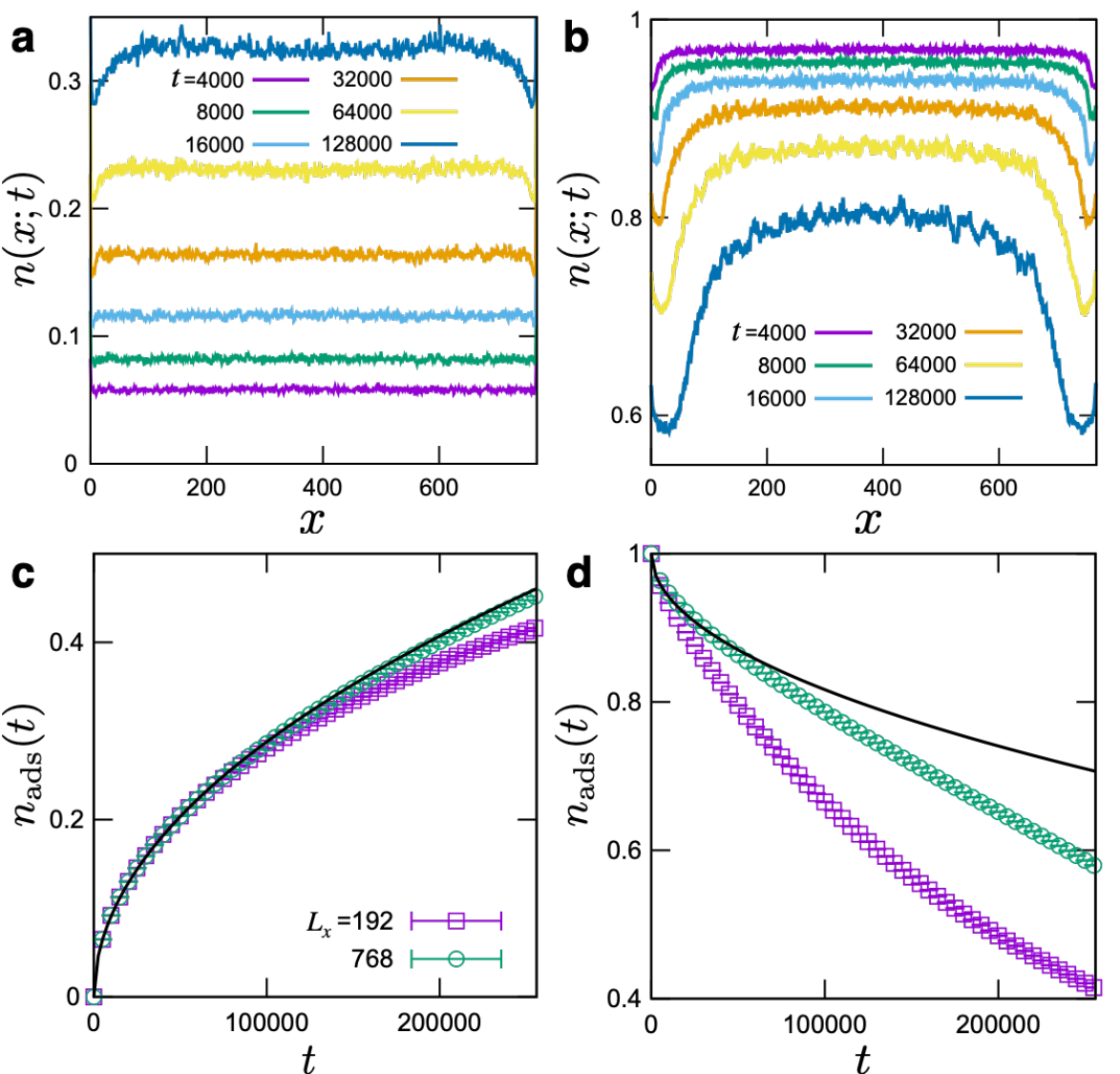}
\caption{
{\bf Adsorption and desorption kinetics for a system with lattice expansion and softening upon adsorption.}
{\bf a,b,} Spatial distribution of the adsorbed particle fraction $n(x;t)$ along the $x$ direction for an adsorption ({\bf a}) and desorption process ({\bf b}). The parameters are $(K,\lambda)=(0.5,1.6)$ so that lattice expansion and softening occur upon adsorption, with $(T,\mu)=(0.08,-0.6)$ in {\bf a} and $(0.08,-1.4)$ in {\bf b}. Data are averaged over 300 independent runs. In the early stage of the adsorption process, the distribution is almost flat because the softer domain favours the flat shape along the boundary (see Extended Data Fig.~\ref{fig::ex::schematic}e). For $t\ge 64000$, the distribution near the corners becomes smaller. This is because softer adsorbed domains are elongated in the bulk due to elastic heterogeneity and reach the top boundary away from the corners to form domains. In the desorption process, on the other hand, desorption proceeds from the corner region even in the early stage (see Extended Data Fig.~\ref{fig::ex::schematic}f). Since the desorbed sites are more rigid, stress relaxation is more likely to occur at the corners than in the surface region, resulting in the enhancement of desorption at the corners.
{\bf c,d,} Time evolution of the total adsorption fraction $n_{\rm ads}(t)$ for the adsorption ({\bf c}) and desorption processes ({\bf d}). The solid curve represents a reference for diffusive growth proportional to $t^{1/2}$. In the adsorption process, the adsorption fraction is suppressed below $t^{1/2}$ due to slow corner adsorption. In the desorption process, on the other hand, fast desorption at the corners results in enhanced desorption faster than $t^{1/2}$. Both in {\bf c} and {\bf d}, the corner effect becomes more conspicuous as the lateral system length $L_x$ decreases, resulting in more pronounced suppression and enhancement of adsorption and desorption, respectively. The system size is $L_x=768$ in panels {\bf a,b}.
}
\label{fig::ex::Fig_dist_uptake_soft}
\end{figure*}

%7
\begin{figure*}[t]
\centering
\includegraphics[width=140mm]{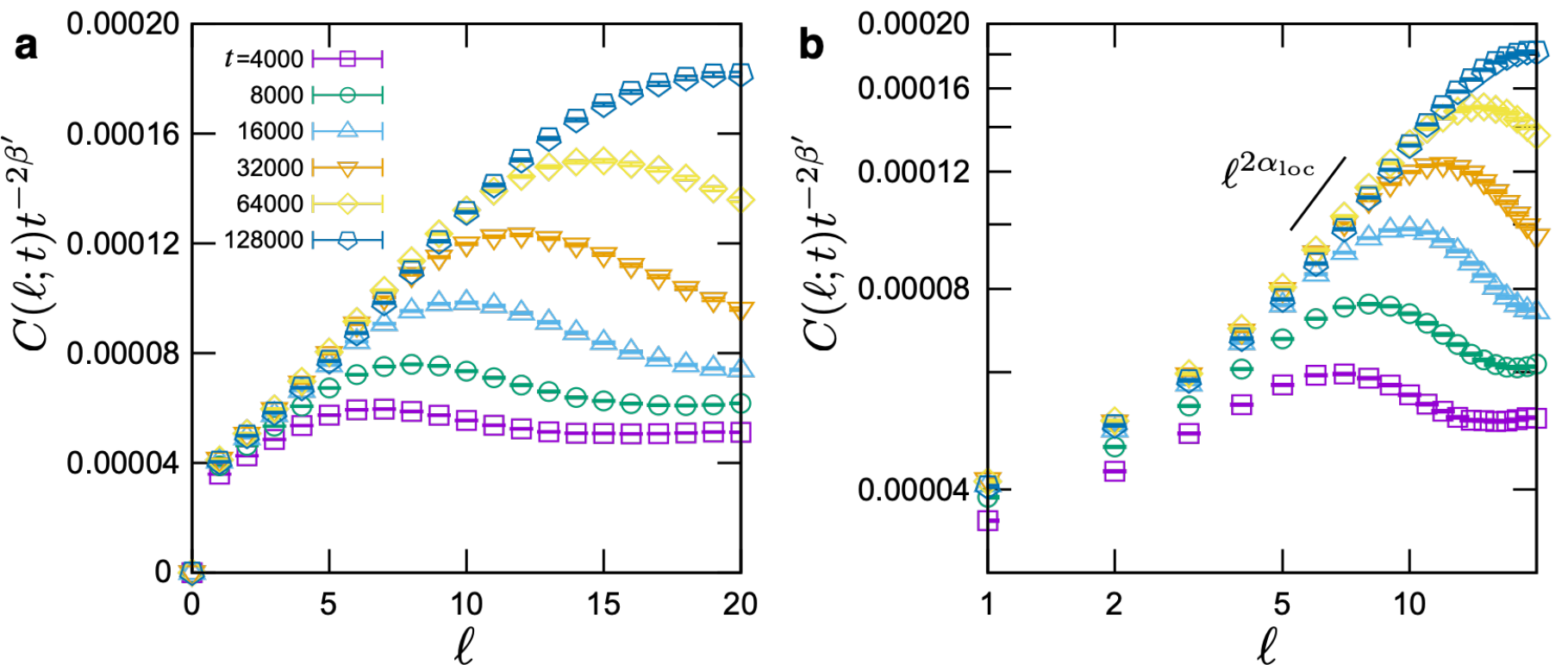}
\caption{
{\bf Small length scale correlation function.}
{\bf a,b,} The scaling plot of the spatial correlation function $C(\ell;t)$ in Eq.(\ref{eq::C}) for small length scales ({\bf a}) and its log-log plot ({\bf b}), scaled by $t^{-2\beta'}$ with $\beta'=0.250$ being the anomalous growth exponent obtained in Fig.~\ref{fig::scaling}b. The data are well-collapsed up to $\ell\lesssim\ell_{\rm peak}(t)$, at which $C(\ell;t)$ exhibits a peak. In {\bf b}, $C(\ell;t)t^{-2\beta'}\sim\ell^{2\alpha_{\rm loc}}$ is confirmed for $5<\ell\lesssim\ell_{\rm peak}$, where the local roughness exponent $\alpha_{\rm loc}=0.424<1$ is obtained in Fig.~\ref{fig::scaling}f. Thus, $C(\ell;t)$ exhibits a feature of intrinsic anomalous scaling for $5<\ell\lesssim\ell_{\rm peak}$. However, the slope becomes more gentle for $\ell<5$. This is due to the discreteness of the numerical model, while the scaling theory is derived from continuum models. The discreteness is also responsible for the deviation from the scaling of the structure factor, as shown in Fig.~\ref{fig::scaling}i. The system size is $L_x=768$.
}
\label{fig::ex::correlation_supple}
\end{figure*}

%8
\begin{figure*}[t]
\centering
\includegraphics[width=160mm]{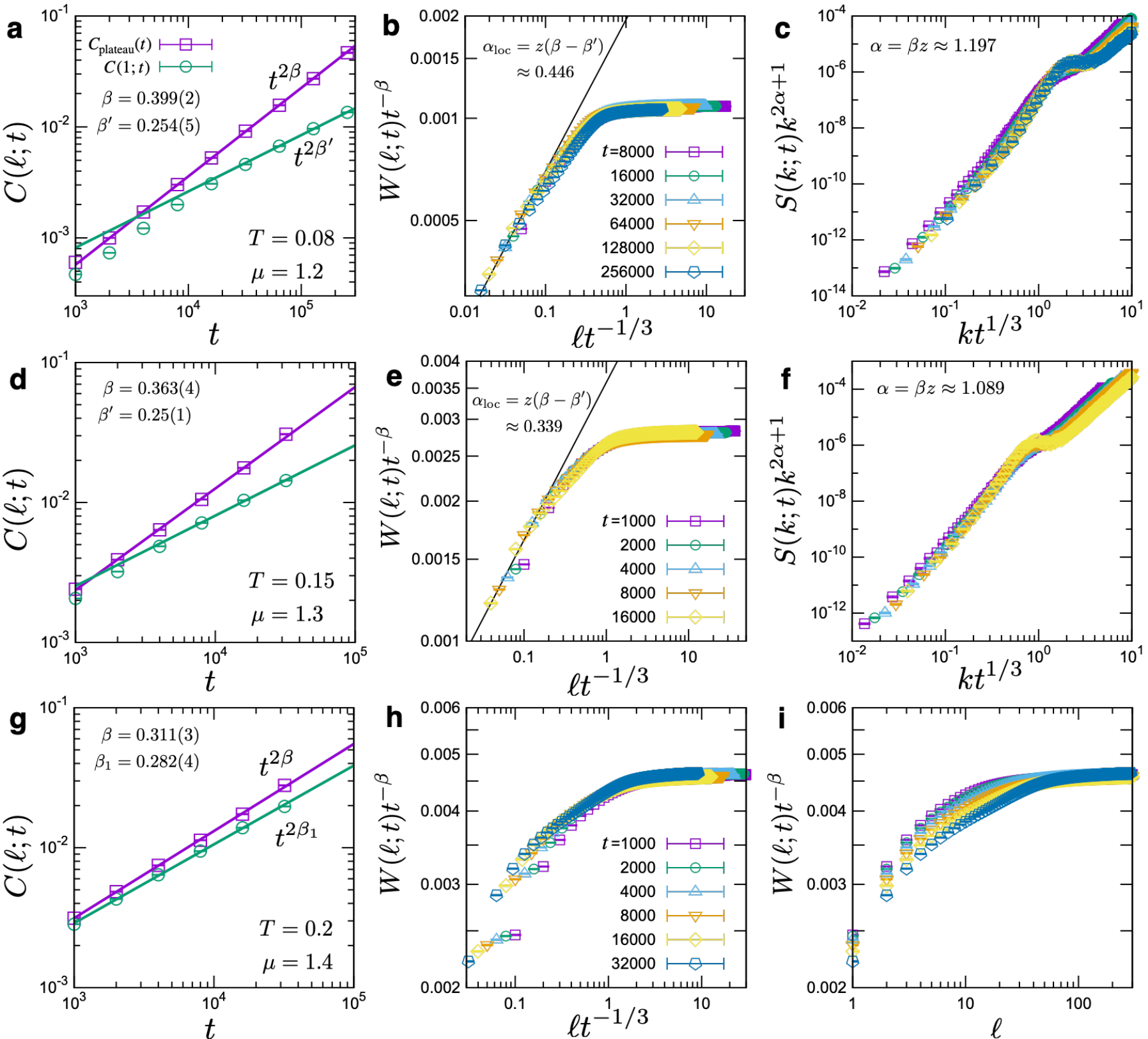}
\caption{
{\bf Dynamic scaling for adsorption kinetics in the spinodal and crossover regions.}
{\bf a-i,} Power law behaviours of the spatial correlation function $C(\ell;t)$ in Eq.(\ref{eq::C}), dynamic scaling of the adsorption deviation $W(\ell;t)$ in Eq.(\ref{eq::W}), and the structure factor $S(k;t)$ in Eq.(\ref{eq::S}) at different thermodynamic parameters. The analysis is performed for two parameter sets in the spinodal region $(T,\mu)=(0.08,1.2)$ ({\bf a-c}) and $(T,\mu)=(0.15,1.3)$ ({\bf d-f}), and for one set in the crossover region, $(T,\mu)=(0.20,1.4)$ ({\bf g-i}). The location of these points is shown in the phase diagram in Fig.~\ref{fig::summary}c. In the spinodal region, the adsorption kinetics exhibits anomalous dynamic scaling. As shown in {\bf a} and {\bf d}, $C_{\rm plateau}(t)\sim t^{2\beta}$ and $C(1;t)\sim t^{2\beta'}$ hold, where $\beta$ and $\beta'$ denote the growth exponent and the anomalous growth exponent. The condition $\beta'>0$ with finite $1/z$ indicates the existence of anomalous scaling. In the dynamic scaling analysis of $W(\ell;t)$ in {\bf b} and {\bf e}, the horizontal and vertical axes are scaled by $t^{-1/3}$ and $t^{-\beta}$, corresponding to the dynamic exponent $1/z=1/3$. For $\ell t^{-1/3} \ll 1$, the slope shows good agreement with the local roughness exponent $\alpha_{\rm loc}=z(\beta-\beta') < 1$, suggesting that the dynamics does not belong to superroughening. Moreover, in the dynamic scaling analysis of $S(k;t)$ in {\bf c} and {\bf f}, the horizontal and vertical axes are scaled by $t^{1/3}$ and $k^{2\alpha+1}$, where $\alpha = \beta z$ denotes the global roughness exponent. A plateau is observed around $kt^{1/3} \simeq 1$, indicating the spectral roughness exponent $\alpha_{\rm s}$ equals $\alpha$. Thus, the dynamic scaling is anomalous but does not belong to either intrinsic anomalous scaling or superroughening. In the crossover region {\bf g-i}, the kinetics do not exhibit any dynamic scaling. In {\bf g}, both $C_{\rm plateau}(t)$ and $C(1;t)$ exhibit power laws with $2\beta$ and $2\beta_1$, respectively. However, $W(\ell;t)$ cannot be scaled by any dynamic exponent $z$. Dynamic scaling plots of $W(\ell;t)$ using $1/z=1/3$ ({\bf h}) and $1/z=0$ ({\bf i}) do not exhibit data collapse, implying there is no characteristic length scale. Thus, the exponent $\beta_1$ is distinguished from the anomalous growth exponent $\beta'$. The roughness exponent $\alpha=\beta z$ cannot be well-defined in this region. The system size for all panels is $L_x=768$.
}
\label{fig::ex::family_vicsek_ex}
\end{figure*}

%9
\begin{figure*}[t]
\centering
\includegraphics[width=160mm]{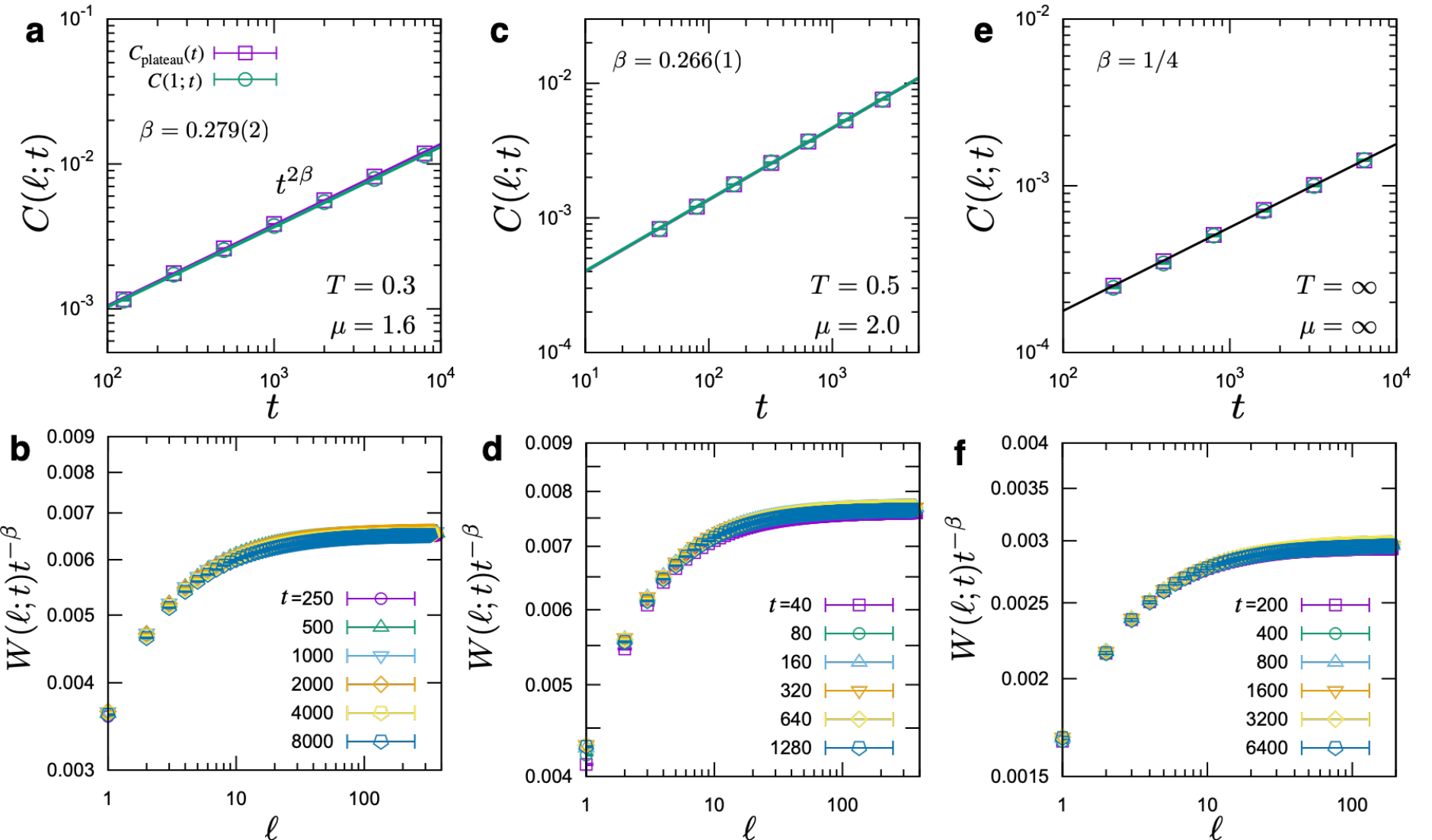}
\caption{
{\bf Dynamic scaling for adsorption kinetics in the supercritical region.}
{\bf a-f,} Power law behaviours of spatial correlation function $C(\ell;t)$ in Eq.(\ref{eq::C}) and dynamic scaling of adsorption deviation $W(\ell;t)$ in Eq.(\ref{eq::W}) in the supercritical region. The analysis is performed for $(T,\mu)=(0.30,1.6)$ ({\bf a, b}), $(T,\mu)=(0.50,2.0)$ ({\bf c, d}), and $(T,\mu)=(\infty,\infty)$ ({\bf e, f}). Both $C_{\rm plateau}(t)$ and $C(1;t)$ exhibit the same power law $t^{2\beta}$, where $\beta$ denotes the growth exponent. This is because the correlation vanishes even at $\ell=1$ for all times, indicating the absence of a characteristic length scale. Correspondingly, dynamic scaling plots of $W(\ell;t)$ are well-collapsed by adopting the dynamic exponent $1/z=0$. The system size for all panels is $L_x=768$.
}
\label{fig::ex::family_vicsek_ex2}
\end{figure*}

%10
\begin{figure*}[t]
\centering
\includegraphics[width=160mm]{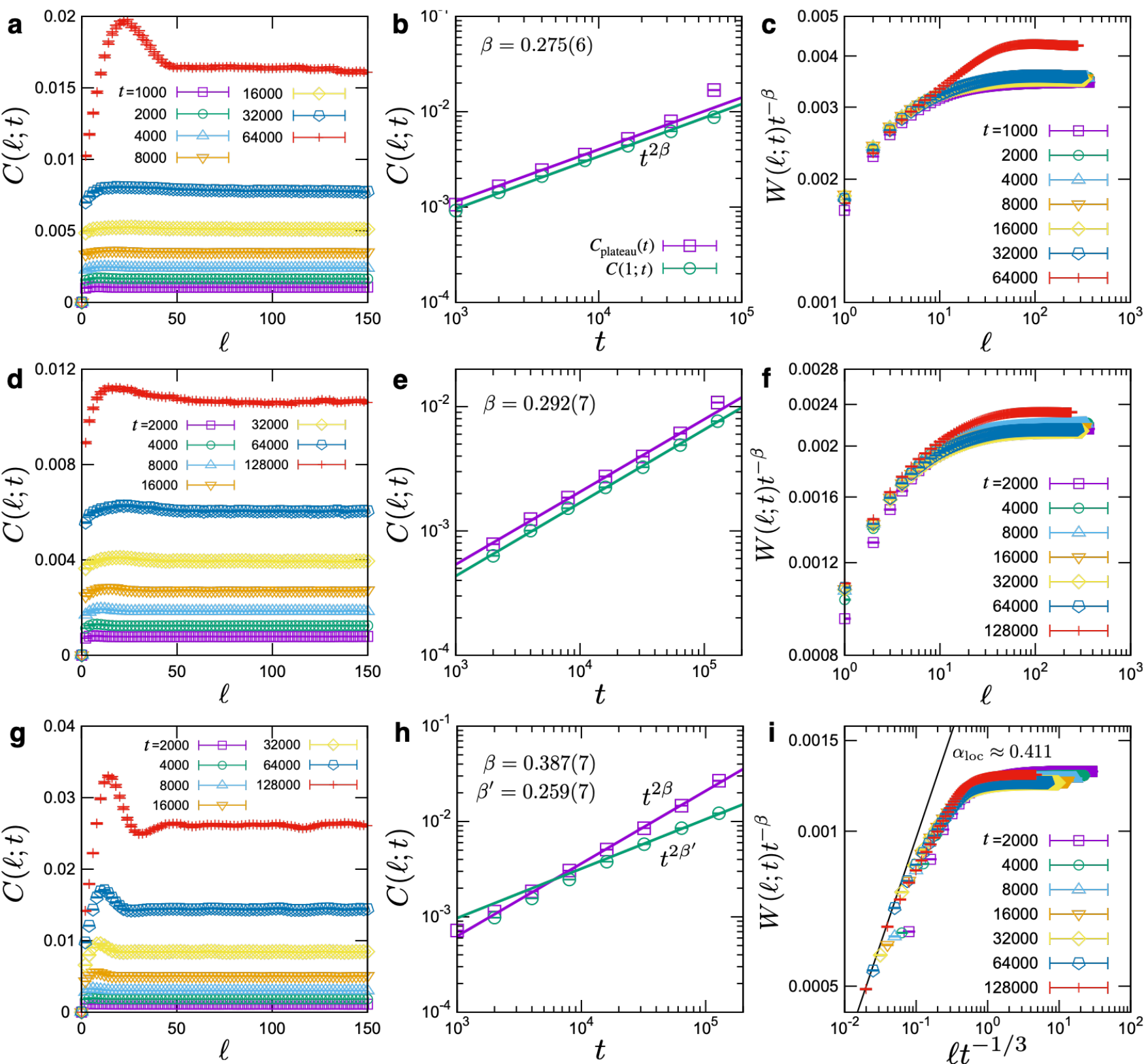}
\caption{
{\bf Dynamic scaling in various situations.}
{\bf a-c,} Desorption kinetics for a system with guest-induced lattice hardening and expansion $(K,\lambda)=(3,1.4)$ at $(T,\mu)=(0.1,0.3)$. The panels show the spatial correlation function $C(\ell;t)$ in Eq.(\ref{eq::C}) ({\bf a}), the plateau height $C_{\rm plateau}(t)$ and the shortest correlation $C(1;t)$ ({\bf b}), and the dynamic scaling of the adsorption deviation $W(\ell;t)$ in Eq.(\ref{eq::W}) ({\bf c}). $C(\ell;t)$ does not have a prominent peak for $t\le 32000$, implying the absence of a growing correlation length $\xi$; thus, the dynamic exponent is $1/z=0$. The desorbed domain observed in Extended Data Fig.~\ref{fig::ex::schematic}d is more flexible, resulting in a uniform distribution in the lateral dimension. Consequently, $C_{\rm plateau}(t)$ and $C(1;t)$ follows the same power law $t^{2\beta}$, and $W(\ell;t)$ exhibits the dynamic scaling $W(\ell;t)=t^\beta\tilde{W}_2(\ell t^{-1/z})$ with $1/z=0$, where $\beta$ denotes the growth exponent. For $t=64000$, on the other hand, a peak structure appears. Narrow desorbed channels penetrate into the adsorbed matrix, as shown in Extended Data Fig.~\ref{fig::ex::schematic}d. Thus, $W(\ell;t)$ at $t=64000$ deviates from the scaling, as shown in {\bf c}.
{\bf d-f,} Adsorption kinetics for a system with guest-induced lattice softening and expansion $(K,\lambda)=(0.5,1.6)$ at $(T,\mu)=(0.08,-0.6)$. Similar to {\bf a-c}, $C(\ell;t)$ does not have a prominent peak for $t\le64000$ because the growing domain is softer also in this case (see Extended Data Fig.~\ref{fig::ex::schematic}e). The same scaling also holds for $t\le 64000$. 
{\bf g-i,} Desorption kinetics for a system with guest-induced lattice softening and expansion $(K,\lambda)=(0.5,1.6)$ at $(T,\mu)=(0.08,-1.4)$. In this case, the growing domains are more rigid, similar to those in Fig.~\ref{fig::scaling} in the main text. $C(\ell;t)$ has a peak even in the early stage, and the peak position shifts to larger $\ell$ as time proceeds. In {\bf i}, dynamic scaling is satisfied with the dynamic exponent being $1/z=1/3$. The roughness exponent $\alpha=\beta z\approx 1.16$ indicates that anomalous roughening occurs. This corresponds to the undulating pattern at the surface, as shown in Extended Data Fig.~\ref{fig::ex::schematic}f. The system size is $L_x=768$.
}
\label{fig::ex::family_vicsek_other}
\end{figure*}

\end{document}